\documentstyle[12pt]{article}
\begin{document}

\author{C. Bizdadea\thanks{%
e-mail addresses: bizdadea@central.ucv.ro and bizdadea@hotmail.com}, E. M.
Cioroianu\thanks{%
e-mail address: manache@central.ucv.ro}, S. O. Saliu\thanks{%
e-mail addresses: osaliu@central.ucv.ro} \\
Department of Physics, University of Craiova\\
13 A. I. Cuza Str., Craiova RO-1100, Romania}
\title{Irreducible Hamiltonian BRST-anti-BRST symmetry for reducible systems}
\maketitle

\begin{abstract}
An irreducible Hamiltonian BRST-anti-BRST treatment of reducible first-class
systems based on homological arguments is proposed. The general formalism is
exemplified on the Freedman-Townsend model.

\end{abstract}

\section{Introduction}

The BRST formalism has been extended over the last few years to a more
symmetrical approach, called the BRST-anti-BRST method. This type of
extended BRST symmetry has been implemented at the Hamiltonian \cite{1}--%
\cite{6}, as well as at the Lagrangian level \cite{6}--\cite{19}. Although
it does not play such an important role like the BRST symmetry itself, the
BRST-anti-BRST symmetry is, however, a useful tool in the geometric
(superfield) description of the BRST transformation, for the perturbative
investigation of the renormalizability of Yang-Mills theories, in a
consistent approach to anomalies or for a correct understanding of the
non-minimal sector from the BRST treatment \cite{20}--\cite{26}. The
BRST-anti-BRST method can be equally used to investigate irreducible and
reducible gauge theories or Hamiltonian systems possessing irreducible and
reducible first-class constraints. Nevertheless, the Hamiltonian
BRST-anti-BRST treatment of reducible first-class systems turns out to be a
complicated mechanism, due, in principal, to the additional redundancy of
the constraint functions implied by the inner structure of the
BRST-anti-BRST symmetry. This makes the computation of the total BRST charge
and BRST-anti-BRST-invariant Hamiltonian, and especially of the gauge-fixed
action, a more difficult task than in the irreducible case.

In this paper we give an irreducible Hamiltonian BRST-anti-BRST treatment of
on-shell reducible first-class theories that somehow simplifies the standard
reducible approach. Our treatment is mainly based on the following steps:
(a) we transform the original reducible first-class constraints into some
irreducible ones on a larger phase-space in a manner that allows the
substitution of the Hamiltonian BRST-anti-BRST symmetry of the reducible
system by that of the irreducible theory; (b) we quantize the irreducible
system accordingly with the Hamiltonian BRST-anti-BRST formalism. This
programme will lead to the removal of all ghosts for ghosts and antighosts
associated with the original reducibility. The idea of replacing a redundant
theory with an irreducible one is not new \cite{1}, \cite{27}, and has
recently been developed at the antifield BRST level for both Lagrangian
gauge theories and Hamiltonian first-class systems \cite{28}. The same idea
has also been approached within the framework of the Lagrangian
BRST-anti-BRST context \cite{29}, but it has neither been consistently
developed, nor yet applied to the Hamiltonian BRST-anti-BRST approach to
on-shell reducible first-class systems, hence our paper establishes a new
result.

The paper is organized into six sections. Section 2 realizes a brief review
of the Hamiltonian BRST-anti-BRST formalism for reducible first-class
systems. Section 3 is focused on the derivation of an irreducible
Koszul-Tate bicomplex associated with a reducible theory. This Koszul-Tate
bicomplex is inferred from the requirement that all nontrivial co-cycles at
total resolution degree one from the reducible approach exclusively due to
the original redundancy become trivial under a proper redefinition of the
antighosts with total resolution degree equal to one. This will underline an
irreducible first-class constraint set corresponding to the initial
reducible one. Section 4 is concerned with the construction of the
Hamiltonian BRST-anti-BRST symmetry of the irreducible theory. In the
meantime, we establish the link with the reducible BRST-anti-BRST symmetry
and show that it is permissible to replace the Hamiltonian BRST-anti-BRST
symmetry of the initial reducible system with that of the irreducible
theory. In section 5 we apply the theoretical part of the paper to the
Freedman-Townsend model. Section 6 ends the paper with the main conclusions.

\section{Hamiltonian BRST-anti-BRST symmetry for reducible first-class
systems}

In this section we give a brief review of the Hamiltonian BRST-anti-BRST
symmetry for reducible first-class systems. Our starting point is a
Hamiltonian system on a phase-space described locally by $N$ bosonic
co-ordinates $z^{A}$, subject to the first-class constraints 
\begin{equation}
\Sigma :G_{a_{0}}\left( z^{A}\right) \approx 0,\;a_{0}=1,\cdots ,M_{0},
\label{1}
\end{equation}
which are assumed to be on-shell $L$-stage reducible. We presume that the
second-class constraints (if any) have been eliminated by means of the Dirac
bracket. The first-class behaviour of the constraints is translated into 
\begin{equation}
\left[ G_{a_{0}},G_{b_{0}}\right] =C_{\;\;a_{0}b_{0}}^{c_{0}}G_{c_{0}},
\label{2}
\end{equation}
while the on-shell redundancy is written as 
\begin{equation}
Z_{\;\;a_{1}}^{a_{0}}G_{a_{0}}=0,\;a_{1}=1,\cdots ,M_{1},  \label{3}
\end{equation}
\begin{equation}
Z_{\;\;a_{2}}^{a_{1}}Z_{\;\;a_{1}}^{a_{0}}=C_{a_{2}}^{a_{0}b_{0}}G_{b_{0}},%
\;a_{2}=1,\cdots ,M_{2},  \label{4}
\end{equation}
\[
\vdots 
\]
\begin{equation}
Z_{\;\;a_{L}}^{a_{L-1}}Z_{\;%
\;a_{L-1}}^{a_{L-2}}=C_{a_{L}}^{a_{L-2}b_{0}}G_{b_{0}},\;a_{L}=1,\cdots
,M_{L},  \label{5}
\end{equation}
where the symbol $\left[ ,\right] $ denotes either the Poisson bracket, or
the Dirac bracket if any second-class constraints were present. All the
functions in (\ref{2}--\ref{5}) may involve the phase-space coordinates,
while the coefficients $C_{a_{2}}^{a_{0}b_{0}}$ and the first-order
structure functions $C_{\;\;a_{0}b_{0}}^{c_{0}}$ are antisymmetric in the
upper, respectively, lower indices. The analysis is performed in the bosonic
case, but can be easily extended to fermions by introducing some appropriate
sign factors.

The Hamiltonian BRST-anti-BRST symmetry for such a reducible system is given
by two anticommuting differentials 
\begin{equation}
s_{Ra}s_{Rb}+s_{Rb}s_{Ra}=0,\;a,b=1,2,  \label{6}
\end{equation}
that can be made to split as 
\begin{equation}
s_{Ra}=\delta _{Ra}+D_{Ra}+\cdots ,\;a=1,2.  \label{7}
\end{equation}
The operators $\left( \delta _{Ra}\right) _{a=1,2}$ generate the Koszul-Tate
differential bicomplex, which is bigraded accordingly the resolution
bidegree $bires=\left( res_{1},res_{2}\right) $, and furnishes an
homological biresolution of smooth functions defined on the first-class
surface (\ref{1}), $C^{\infty }\left( \Sigma \right) $. $\left(
D_{Ra}\right) _{a=1,2}$ are known as longitudinal exterior derivatives and
provide the extended longitudinal bicomplex bigraded via the form bidegree $%
biform=\left( form_{1},form_{2}\right) $, which offers a redundant
description of the tangent space to the gauge orbits. We set, as usual, $%
bires\left( \delta _{R1}\right) =(-1,0)$, $bires\left( \delta _{R2}\right)
=(0,-1)$, $biform\left( D_{R1}\right) =(1,0)$, $biform\left( D_{R2}\right)
=(0,1)$, $bires\left( D_{R1}\right) =(0,0)$, $bires\left( D_{R2}\right)
=(0,0)$. The remaining terms in $\left( s_{Ra}\right) _{a=1,2}$, generically
denoted by ``$\cdots $'', are required in order to ensure the BRST-anti-BRST
algebra defining relations (\ref{6}). The generators of the Koszul-Tate
bicomplex are known as antighosts, while those from the exterior
longitudinal bicomplex are called ghosts. The operators $s_{R1}$ and $s_{R2}$
constitute the basic ingredients of the BRST-anti-BRST differential
bicomplex, with the bigrading given by the new ghost bidegree $bingh=\left(
ngh_{1},ngh_{2}\right) $ defined like 
\begin{equation}
\left( ngh_{1},ngh_{2}\right) =\left(
form_{1}-res_{1},form_{2}-res_{2}\right) ,  \label{8}
\end{equation}
which is such that $bingh\left( s_{R1}\right) =\left( 1,0\right) $ and $%
bingh\left( s_{R2}\right) =\left( 0,1\right) $. Consequently, we have that $%
bingh\left( \delta _{R1}\right) =\left( 1,0\right) $, $bingh\left( \delta
_{R2}\right) =\left( 0,1\right) $, $bingh\left( D_{R1}\right) =\left(
1,0\right) $, $bingh\left( D_{R2}\right) =\left( 0,1\right) $. The crucial
property of this double complex is that the zeroth order cohomology spaces
of $s_{R1}$ and $s_{R2}$ are isomorphic to the algebra of physical
observables (the algebra of gauge invariant functions defined on (\ref{1})).
Finally, a word on the relationship between the BRST-anti-BRST and BRST
symmetries. It has been shown \cite{5} that the Hamiltonian BRST-anti-BRST
symmetry for an arbitrary Hamiltonian first-class system exists provided the
standard Hamiltonian BRST symmetry for the system can be properly
constructed. Actually, the sum between the BRST and anti-BRST operators
(total BRST transformation) 
\begin{equation}
s_{R}=s_{R1}+s_{R2},  \label{9}
\end{equation}
defines a simple BRST complex associated with a complete description of the
first-class surface (\ref{1}) obtained by doubling the first-class
constraints, which is graded accordingly as the new ghost number $%
ngh=ngh_{1}+ngh_{2}$ ($ngh\left( s_{R}\right) =1$). Accordingly, we have
that the total BRST symmetry splits in the usual way as 
\begin{equation}
s_{R}=\delta _{R}+D_{R}+\cdots ,  \label{10}
\end{equation}
where 
\begin{equation}
\delta _{R}=\delta _{R1}+\delta _{R2},  \label{11}
\end{equation}
gives a simple Koszul-Tate complex graded by the total resolution degree $%
res=res_{1}+res_{2}$ ($res\left( \delta _{R}\right) =-1$), that realizes a
homological resolution of smooth functions defined on (\ref{1})
corresponding to the redundant description of this first-class surface, the
operator 
\begin{equation}
D_{R}=D_{R1}+D_{R2},  \label{12}
\end{equation}
leads to a simple exterior longitudinal complex graded by the total form
degree $form=form_{1}+form_{2}$ ($form\left( D_{R}\right) =1$, $res\left(
D_{R}\right) =0$), which offers a redundant description of the tangent space
to the gauge orbits, and the rest of the terms ``$\cdots $'' ensure the
nilpotency of $s_{R}$%
\begin{equation}
s_{R}^{2}=0.  \label{13}
\end{equation}
Of course, the zeroth order cohomology space of the total BRST
transformation is again isomorphic to the algebra of physical observables 
\begin{equation}
H^{0}\left( s_{R}\right) \simeq \left\{ {\rm physical\;observables}\right\} .
\label{14}
\end{equation}
The link between the new ghost bidegree and the usual ghost number ($gh$)
from the standard Hamiltonian BRST formalism is expressed by $%
gh=ngh_{1}-ngh_{2}$.

After this brief review on the main ideas underlying the Hamiltonian
BRST-anti-BRST formalism, we analyse the basic steps in the construction of
the Koszul-Tate bicomplex, which should be performed in such a way to comply
with the essential requirements, which are the nilpotency and
anticommutivity of $\left( \delta _{Ra}\right) _{a=1,2}$ 
\begin{equation}
\delta _{Ra}\delta _{Rb}+\delta _{Rb}\delta _{Ra}=0,\;a,b=1,2,  \label{14a}
\end{equation}
together with the conditions that define the biresolution property 
\begin{equation}
H_{0,0}\left( \delta _{Ra}\right) =C^{\infty }\left( \Sigma \right) ,\;a=1,2,
\label{14b}
\end{equation}
\begin{equation}
H_{j,k}\left( \delta _{Ra}\right) =0,\;j,k\geq 0,\;j+k\neq 0,\;a=1,2,
\label{14c}
\end{equation}
where $H_{j,k}\left( \delta _{Ra}\right) $ denotes the space of elements
with the resolution bidegree equal to $(j,k)$ that are $\delta _{Ra}$-closed
modulo $\delta _{Ra}$-exact (in the following we will suggestively refer to (%
\ref{14c}) as the `biacyclicity conditions'). In the case of a first-stage
reducible Hamiltonian system ($L=1$), subject to the first-class constraints
(\ref{1}) and the reducibility relations (\ref{3}), we introduce the
antighost spectrum \cite{5} 
\begin{equation}
\left( \stackrel{\lbrack 1,0]}{\cal P}_{1a_{0}},\stackrel{[0,1]}{\cal P}%
_{2a_{0}}\right) ,\;\left( \stackrel{[2,0]}{P}_{1a_{1}},\stackrel{[1,1]}{P}%
_{2a_{1}},\stackrel{[0,2]}{P}_{3a_{1}},\stackrel{[1,1]}{\lambda }%
_{a_{0}}\right) ,  \label{15}
\end{equation}
\begin{equation}
\left( \stackrel{\lbrack 2,1]}{\rho }_{1a_{1}},\stackrel{[1,2]}{\rho }%
_{2a_{1}}\right) ,  \label{15a}
\end{equation}
where $\stackrel{[m,n]}{F}$ signifies an object with $bires\left( F\right)
=(m,n)$. The relation between $bingh$ and $bires$ in the case of all
variables from the antighost spectrum is $bingh=\left(
-res_{1},-res_{2}\right) $. The proper definitions of $\left( \delta
_{Ra}\right) _{a=1,2}$ acting on the generators from the Koszul-Tate
bicomplex are given by 
\begin{equation}
\delta _{R1}\stackrel{[0,0]}{z}^{A}=0,\;\delta _{R2}\stackrel{[0,0]}{z}%
^{A}=0,  \label{16}
\end{equation}
\begin{equation}
\delta _{R1}\stackrel{[1,0]}{\cal P}_{1a_{0}}=-G_{a_{0}},\;\delta _{R2}%
\stackrel{[1,0]}{\cal P}_{1a_{0}}=0,  \label{17}
\end{equation}
\begin{equation}
\delta _{R1}\stackrel{[0,1]}{\cal P}_{2a_{0}}=0,\;\delta _{R2}\stackrel{[0,1]%
}{\cal P}_{2a_{0}}=-G_{a_{0}},  \label{18}
\end{equation}
\begin{equation}
\delta _{R1}\stackrel{[2,0]}{P}_{1a_{1}}=-\stackrel{[1,0]}{\cal P}%
_{1a_{0}}Z_{\;\;a_{1}}^{a_{0}},\;\delta _{R2}\stackrel{[2,0]}{P}_{1a_{1}}=0,
\label{19}
\end{equation}
\begin{equation}
\delta _{R1}\stackrel{[1,1]}{P}_{2a_{1}}=\frac{1}{2}\stackrel{[0,1]}{\cal P}%
_{2a_{0}}Z_{\;\;a_{1}}^{a_{0}},\;\delta _{R2}\stackrel{[1,1]}{P}_{2a_{1}}=%
\frac{1}{2}\stackrel{[1,0]}{\cal P}_{1a_{0}}Z_{\;\;a_{1}}^{a_{0}},
\label{20}
\end{equation}
\begin{equation}
\delta _{R1}\stackrel{[0,2]}{P}_{3a_{1}}=0,\;\delta _{R2}\stackrel{[0,2]}{P}%
_{3a_{1}}=-\stackrel{[0,1]}{\cal P}_{2a_{0}}Z_{\;\;a_{1}}^{a_{0}},
\label{21}
\end{equation}
\begin{equation}
\delta _{R1}\stackrel{[1,1]}{\lambda }_{a_{0}}=-\stackrel{[0,1]}{\cal P}%
_{2a_{0}},\;\delta _{R2}\stackrel{[1,1]}{\lambda }_{a_{0}}=\stackrel{[1,0]}%
{\cal P}_{1a_{0}},  \label{22}
\end{equation}
\begin{equation}
\delta _{R1}\stackrel{[2,1]}{\rho }_{1a_{1}}=-\stackrel{[1,1]}{P}_{2a_{1}}-%
\frac{1}{2}\stackrel{[1,1]}{\lambda }_{a_{0}}Z_{\;\;a_{1}}^{a_{0}},\;\delta
_{R2}\stackrel{[2,1]}{\rho }_{1a_{1}}=-\stackrel{[2,0]}{P}_{1a_{1}},
\label{23}
\end{equation}
\begin{equation}
\delta _{R1}\stackrel{[1,2]}{\rho }_{2a_{1}}=-\stackrel{[0,2]}{P}%
_{3a_{1}},\;\delta _{R2}\stackrel{[1,2]}{\rho }_{2a_{1}}=-\stackrel{[1,1]}{P}%
_{2a_{1}}+\frac{1}{2}\stackrel{[1,1]}{\lambda }_{a_{0}}Z_{\;\;a_{1}}^{a_{0}},
\label{24}
\end{equation}
and ensure the basic conditions (\ref{14a}--\ref{14c}). For a second-stage
first-class Hamiltonian theory ($L=2$) described by the constraints (\ref{1}%
) and the reducibility relations (\ref{3}--\ref{4}), one needs to supplement
the antighost spectrum (\ref{15}--\ref{15a}) with the antighosts 
\begin{equation}
\left( \stackrel{\lbrack 3,0]}{P}_{1a_{2}},\stackrel{[2,1]}{P}_{2a_{2}},%
\stackrel{[1,2]}{P}_{3a_{2}},\stackrel{[0,3]}{P}_{4a_{2}}\right) ,\left( 
\stackrel{[3,1]}{\rho }_{1a_{2}},\stackrel{[2,2]}{\rho }_{2a_{2}},\stackrel{%
[1,3]}{\rho }_{3a_{2}}\right) ,  \label{25}
\end{equation}
and set the definitions (\ref{16}--\ref{24}) together with 
\begin{equation}
\delta _{R1}\stackrel{[3,0]}{P}_{1a_{2}}=-\stackrel{[2,0]}{P}%
_{1a_{1}}Z_{\;\;a_{2}}^{a_{1}}+\frac{1}{2}C_{a_{2}}^{a_{0}b_{0}}\stackrel{%
[1,0]}{\cal P}_{1a_{0}}\stackrel{[1,0]}{\cal P}_{1b_{0}},\;\delta _{R2}%
\stackrel{[3,0]}{P}_{1a_{2}}=0,  \label{26}
\end{equation}
\begin{equation}
\delta _{R1}\stackrel{[2,1]}{P}_{2a_{2}}=\frac{1}{2}\stackrel{[1,1]}{P}%
_{2a_{1}}Z_{\;\;a_{2}}^{a_{1}}+\frac{1}{4}C_{a_{2}}^{a_{0}b_{0}}\stackrel{%
[1,0]}{\cal P}_{1a_{0}}\stackrel{[0,1]}{\cal P}_{2b_{0}},  \label{27}
\end{equation}
\begin{equation}
\delta _{R2}\stackrel{[2,1]}{P}_{2a_{2}}=-\frac{1}{2}\stackrel{[2,0]}{P}%
_{1a_{1}}Z_{\;\;a_{2}}^{a_{1}}+\frac{1}{4}C_{a_{2}}^{a_{0}b_{0}}\stackrel{%
[1,0]}{\cal P}_{1a_{0}}\stackrel{[1,0]}{\cal P}_{1b_{0}},  \label{27a}
\end{equation}
\begin{equation}
\delta _{R1}\stackrel{[1,2]}{P}_{3a_{2}}=\frac{1}{2}\stackrel{[0,2]}{P}%
_{3a_{1}}Z_{\;\;a_{2}}^{a_{1}}-\frac{1}{4}C_{a_{2}}^{a_{0}b_{0}}\stackrel{%
[0,1]}{\cal P}_{2a_{0}}\stackrel{[0,1]}{\cal P}_{2b_{0}},  \label{28}
\end{equation}
\begin{equation}
\delta _{R2}\stackrel{[1,2]}{P}_{3a_{2}}=-\frac{1}{2}\stackrel{[1,1]}{P}%
_{2a_{1}}Z_{\;\;a_{2}}^{a_{1}}-\frac{1}{4}C_{a_{2}}^{a_{0}b_{0}}\stackrel{%
[1,0]}{\cal P}_{1a_{0}}\stackrel{[0,1]}{\cal P}_{2b_{0}},  \label{28a}
\end{equation}
\begin{equation}
\delta _{R1}\stackrel{[0,3]}{P}_{4a_{2}}=0,\;\delta _{R2}\stackrel{[0,3]}{P}%
_{4a_{2}}=-\stackrel{[0,2]}{P}_{3a_{1}}Z_{\;\;a_{2}}^{a_{1}}+\frac{1}{2}%
C_{a_{2}}^{a_{0}b_{0}}\stackrel{[0,1]}{\cal P}_{2a_{0}}\stackrel{[0,1]}{\cal %
P}_{2b_{0}},  \label{29}
\end{equation}
\begin{equation}
\delta _{R1}\stackrel{[3,1]}{\rho }_{1a_{2}}=-\stackrel{[2,1]}{P}_{2a_{2}}-%
\frac{1}{2}\stackrel{[2,1]}{\rho }_{1a_{1}}Z_{\;\;a_{2}}^{a_{1}}+\frac{1}{4}%
C_{a_{2}}^{a_{0}b_{0}}\stackrel{[1,1]}{\lambda }_{a_{0}}\stackrel{[1,0]}%
{\cal P}_{1b_{0}},  \label{30}
\end{equation}
\begin{equation}
\delta _{R2}\stackrel{[3,1]}{\rho }_{1a_{2}}=\stackrel{[3,0]}{P}_{1a_{2}},
\label{30a}
\end{equation}
\begin{equation}
\delta _{R1}\stackrel{[2,2]}{\rho }_{2a_{2}}=-\stackrel{[1,2]}{P}_{3a_{2}}-%
\frac{1}{2}\stackrel{[1,2]}{\rho }_{2a_{1}}Z_{\;\;a_{2}}^{a_{1}}-\frac{1}{4}%
C_{a_{2}}^{a_{0}b_{0}}\stackrel{[1,1]}{\lambda }_{a_{0}}\stackrel{[0,1]}%
{\cal P}_{2b_{0}},  \label{31}
\end{equation}
\begin{equation}
\delta _{R2}\stackrel{[2,2]}{\rho }_{2a_{2}}=-\stackrel{[2,1]}{P}_{2a_{2}}+%
\frac{1}{2}\stackrel{[2,1]}{\rho }_{1a_{1}}Z_{\;\;a_{2}}^{a_{1}}-\frac{1}{4}%
C_{a_{2}}^{a_{0}b_{0}}\stackrel{[1,1]}{\lambda }_{a_{0}}\stackrel{[1,0]}%
{\cal P}_{1b_{0}},  \label{31a}
\end{equation}
\begin{equation}
\delta _{R1}\stackrel{[1,3]}{\rho }_{3a_{2}}=-\stackrel{[0,3]}{P}_{4a_{2}},
\label{32a}
\end{equation}
\begin{equation}
\delta _{R2}\stackrel{[1,3]}{\rho }_{3a_{2}}=-\stackrel{[1,2]}{P}_{3a_{2}}+%
\frac{1}{2}\stackrel{[1,2]}{\rho }_{2a_{1}}Z_{\;\;a_{2}}^{a_{1}}+\frac{1}{4}%
C_{a_{2}}^{a_{0}b_{0}}\stackrel{[1,1]}{\lambda }_{a_{0}}\stackrel{[0,1]}%
{\cal P}_{2b_{0}},  \label{32}
\end{equation}
in order to obtain (\ref{14a}--\ref{14c}). Similarly, in the general case of
an $L$-stage reducible Hamiltonian system possessing the first-class
constraints (\ref{1}) and subject to the reducibility relations (\ref{3}--%
\ref{5}), the Koszul-Tate bicomplex includes the generators 
\begin{equation}
\left( \stackrel{\lbrack 1,0]}{\cal P}_{1a_{0}},\stackrel{[0,1]}{\cal P}%
_{2a_{0}}\right) ,  \label{33}
\end{equation}
\begin{equation}
\left( \stackrel{\lbrack 2,0]}{P}_{1a_{1}},\stackrel{[1,1]}{P}_{2a_{1}},%
\stackrel{[0,2]}{P}_{3a_{1}},\stackrel{[1,1]}{\lambda }_{a_{0}}\right) ,
\label{34}
\end{equation}
\begin{equation}
\left( \stackrel{\lbrack k+1,0]}{P}_{1a_{k}},\stackrel{[k,1]}{P}%
_{2a_{k}},\cdots ,\stackrel{[0,k+1]}{P}_{(k+2)a_{k}},\stackrel{[k,1]}{\rho }%
_{1a_{k-1}},\stackrel{[k-1,2]}{\rho }_{2a_{k-1}},\cdots ,\stackrel{[1,k]}{%
\rho }_{ka_{k-1}}\right) ,  \label{35}
\end{equation}
\begin{equation}
\left( \stackrel{\lbrack L+1,1]}{\rho }_{1a_{L}},\stackrel{[L,2]}{\rho }%
_{2a_{L}},\cdots ,\stackrel{[1,L+1]}{\rho }_{(L+1)a_{L}}\right) ,  \label{36}
\end{equation}
with $k=2,\cdots ,L$, on which $\left( \delta _{Ra}\right) _{a=1,2}$ act in
a way that complies with (\ref{14a}--\ref{14c}). We note that the antighosts
(\ref{33}--\ref{36}) are organized on levels of increasing total resolution
degree, all the antighosts in say (\ref{35}) displaying the same resolution
degree, $k+1$.

\section{Irreducible Koszul-Tate bicomplex}

Here, we investigate the possibility of associating an irreducible
Koszul-Tate bicomplex with the initial reducible one. By `irreducible
Koszul-Tate bicomplex' we understand a double complex underlying some
irreducible first-class constraints, and hence whose redundancy is dictated
only by the reducibility relations resulting from the doubling of constraint
functions. The cornerstone of our approach relies on redefining the
antighosts (\ref{33}) of total resolution degree one such that all the
co-cycles of $\left( \delta _{Ra}\right) _{a=1,2}$ at total resolution
degree one due exclusively to the original reducibility relations (\ref{3}--%
\ref{5}) become trivial (either identically vanish or are made exact). Then,
all the corresponding higher-order resolution degree antighosts will be
eliminated from the antighost spectrum, which will retain only the
generators $\stackrel{[1,0]}{\cal P}_{1a_{0}}$, $\stackrel{[0,1]}{\cal P}%
_{2a_{0}}$ and $\stackrel{[1,1]}{\lambda }_{a_{0}}$ from the reducible
treatment. Consequently, we expect the phase-space and antighost spectrum to
be modified in order to furnish a biresolution of smooth functions defined
on the surface of irreducible first-class constraints, but only up to some
new bosonic canonical pairs, respectively, antighosts of resolution
bidegrees $\left( 1,0\right) $, $\left( 0,1\right) $ or $\left( 1,1\right) $%
. For the sake of clarity, we initially analyse the cases $L=1,2$, and then
generalize the results to an arbitrary $L$.

\subsection{First-stage reducible theories}

We begin with a first-stage reducible set of first-class constraints ($L=1$%
), described by the formulae (\ref{1}--\ref{3}). The reducible Koszul-Tate
bicomplex is then fully determined by the generators (\ref{15}--\ref{15a}),
and the definitions (\ref{16}--\ref{24}). The co-cycles of $\left( \delta
_{Ra}\right) _{a=1,2}$ at total resolution degree one caused by the initial
redundancy relations (\ref{3}) are obviously given by 
\begin{equation}
\mu _{a_{1}}^{1}=\stackrel{[1,0]}{\cal P}_{1a_{0}}Z_{\;\;a_{1}}^{a_{0}},\;%
\mu _{a_{1}}^{2}=\stackrel{[0,1]}{\cal P}_{2a_{0}}Z_{\;\;a_{1}}^{a_{0}}.
\label{37}
\end{equation}
Our main concern is to investigate whether or not it is possible to perform
a transformation of the type 
\begin{equation}
\left( \stackrel{\lbrack 1,0]}{\cal P}_{1a_{0}},\stackrel{[0,1]}{\cal P}%
_{2a_{0}}\right) \rightarrow \left( \stackrel{[1,0]}{{\cal P}^{\prime }}%
_{1a_{0}},\stackrel{[0,1]}{{\cal P}^{\prime }}_{2a_{0}}\right) =\left( 
\stackrel{[1,0]}{\cal P}_{1b_{0}}\stackrel{[0,0]}{M}_{\;\;a_{0}}^{b_{0}},%
\stackrel{[0,1]}{\cal P}_{2b_{0}}\stackrel{[0,0]}{N}_{\;\;a_{0}}^{b_{0}}%
\right) ,  \label{38}
\end{equation}
that makes the new co-cycles of the type (\ref{37}) trivial. The above
redefinition of the antighosts is such as to preserve the original
resolution bidegrees. Thus, the matrices $M$ and $N$ can depend at most on
the phase-space variables, $z^{A}$. Taking into account the symmetry between
the BRST and anti-BRST components present everywhere in the development of
the BRST-anti-BRST formalism, it is natural to take 
\begin{equation}
\stackrel{\lbrack 0,0]}{M}_{\;\;a_{0}}^{b_{0}}=\stackrel{[0,0]}{N}%
_{\;\;a_{0}}^{b_{0}}\equiv T_{\;\;a_{0}}^{b_{0}}.  \label{40}
\end{equation}
If we impose on the matrix $T_{\;\;a_{0}}^{b_{0}}$ the conditions 
\begin{equation}
T_{\;\;a_{0}}^{b_{0}}G_{b_{0}}=G_{a_{0}},\;T_{\;\;a_{0}}^{b_{0}}Z_{\;%
\;a_{1}}^{a_{0}}\equiv 0,  \label{41}
\end{equation}
and multiply the former definitions in (\ref{17}) and the latter ones in (%
\ref{18}) by $T_{\;\;a_{0}}^{b_{0}}$, we obtain 
\begin{equation}
\delta _{1}\stackrel{[1,0]}{{\cal P}^{\prime }}_{1a_{0}}=-G_{a_{0}},\;\delta
_{2}\stackrel{[0,1]}{{\cal P}^{\prime }}_{2a_{0}}=-G_{a_{0}}.  \label{40a}
\end{equation}
The formulae (\ref{40a}) will lead to some co-cycles of the type (\ref{37}),
i.e., 
\begin{equation}
\mu _{a_{1}}^{\prime 1}=\stackrel{[1,0]}{{\cal P}^{\prime }}%
_{1a_{0}}Z_{\;\;a_{1}}^{a_{0}},\;\mu _{a_{1}}^{\prime 2}=\stackrel{[0,1]}{%
{\cal P}^{\prime }}_{2a_{0}}Z_{\;\;a_{1}}^{a_{0}},  \label{39}
\end{equation}
that are found to be trivial ($\mu _{a_{1}}^{\prime 1}\equiv 0$, $\mu
_{a_{1}}^{\prime 2}\equiv 0$) thanks to the latter condition in (\ref{41}).
In this way, it results that the co-cycles (\ref{39}) are trivial as long as
equations (\ref{41}) possess solutions. The solution to (\ref{41}) is 
\begin{equation}
T_{\;\;a_{0}}^{b_{0}}=\delta _{\;\;a_{0}}^{b_{0}}-Z_{\;\;b_{1}}^{b_{0}}\bar{D%
}_{\;\;a_{1}}^{b_{1}}A_{a_{0}}^{\;\;a_{1}},  \label{42}
\end{equation}
where the functions $A_{a_{0}}^{\;\;a_{1}}$ can at most involve $z^{A}$ and
are taken to satisfy 
\begin{equation}
rank\left( Z_{\;\;a_{1}}^{a_{0}}A_{a_{0}}^{\;\;b_{1}}\right) \equiv
rank\left( D_{\;\;a_{1}}^{b_{1}}\right) =M_{1},  \label{43}
\end{equation}
with $\bar{D}_{\;\;a_{1}}^{b_{1}}$ denoting the inverse of $%
D_{\;\;a_{1}}^{b_{1}}=Z_{\;\;a_{1}}^{a_{0}}A_{a_{0}}^{\;\;b_{1}}$.
Furthermore, as $\mu _{a_{1}}^{\prime 1}$ and $\mu _{a_{1}}^{\prime 2}$ are
no longer co-cycles, the antighosts $\stackrel{[2,0]}{P}_{1a_{1}}$, $%
\stackrel{[1,1]}{P}_{2a_{1}}$, $\stackrel{[0,2]}{P}_{3a_{1}}$ will be
discarded from the antighost spectrum (\ref{15}), which consequently implies
the removal of $\stackrel{[2,1]}{\rho }_{1a_{1}}$, $\stackrel{[1,2]}{\rho }%
_{2a_{1}}$ from (\ref{15a}). In order to outline the irreducibility of the
resulting Koszul-Tate bicomplex, in (\ref{40a}) we employed the notation $%
\delta _{a}$ instead of $\delta _{Ra}$. Obviously, the complete actions of
the BRST and anti-BRST Koszul-Tate operators on the new generators $%
\stackrel{[1,0]}{{\cal P}^{\prime }}_{1a_{0}}$ and $\stackrel{[0,1]}{{\cal P}%
^{\prime }}_{2a_{0}}$ are realized as 
\begin{equation}
\delta _{1}\stackrel{[1,0]}{{\cal P}^{\prime }}_{1a_{0}}=-G_{a_{0}},\;\delta
_{2}\stackrel{[1,0]}{{\cal P}^{\prime }}_{1a_{0}}=0,  \label{45}
\end{equation}
\begin{equation}
\delta _{1}\stackrel{[0,1]}{{\cal P}^{\prime }}_{2a_{0}}=0,\;\delta _{2}%
\stackrel{[0,1]}{{\cal P}^{\prime }}_{2a_{0}}=-G_{a_{0}}.  \label{46}
\end{equation}
Inserting the relations (\ref{38}) and (\ref{42}) in formulae (\ref{45}--\ref
{46}), we deduce that (\ref{45}--\ref{46}) are equivalent to 
\begin{equation}
\delta _{1}\stackrel{[1,0]}{\cal P}_{1a_{0}}=-G_{a_{0}}+\delta _{1}\left(
A_{a_{0}}^{\;\;a_{1}}\bar{D}_{\;\;a_{1}}^{b_{1}}Z_{\;\;b_{1}}^{b_{0}}%
\stackrel{[1,0]}{\cal P}_{1b_{0}}\right) ,\;\delta _{2}\stackrel{[1,0]}{\cal %
P}_{1a_{0}}=0,  \label{47}
\end{equation}
\begin{equation}
\delta _{1}\stackrel{[0,1]}{\cal P}_{2a_{0}}=0,\;\delta _{2}\stackrel{[0,1]}%
{\cal P}_{2a_{0}}=-G_{a_{0}}+\delta _{2}\left( A_{a_{0}}^{\;\;a_{1}}\bar{D}%
_{\;\;a_{1}}^{b_{1}}Z_{\;\;b_{1}}^{b_{0}}\stackrel{[0,1]}{\cal P}%
_{2b_{0}}\right) .  \label{48}
\end{equation}
Our next concern is to determine the structure of the irreducible
first-class constraints constituting the core of this irreducible bicomplex.
In this light, we enlarge the phase-space by adding some new bosonic
canonical pairs $z^{A_{1}}\equiv \left( y^{a_{1}},\pi _{a_{1}}\right) $ with
resolution bidegree $\left( 0,0\right) $, whose momenta are restricted to be
nonvanishing solutions to the equations 
\begin{equation}
D_{\;\;a_{1}}^{b_{1}}\pi _{b_{1}}=\delta _{1}\left( -Z_{\;\;a_{1}}^{b_{0}}%
\stackrel{[1,0]}{\cal P}_{1b_{0}}\right) ,  \label{49}
\end{equation}
\begin{equation}
D_{\;\;a_{1}}^{b_{1}}\pi _{b_{1}}=\delta _{2}\left( -Z_{\;\;a_{1}}^{b_{0}}%
\stackrel{[0,1]}{\cal P}_{2b_{0}}\right) ,  \label{50}
\end{equation}
and on which $\left( \delta _{a}\right) _{a=1,2}$ are set to act like 
\begin{equation}
\delta _{1}z^{A_{1}}=0,\;\delta _{2}z^{A_{1}}=0.  \label{51}
\end{equation}
On account of the invertibility of $D_{\;\;a_{1}}^{b_{1}}$, the nonvanishing
solutions to the equations (\ref{49}--\ref{50}) ensure the irreducibility as
these equations possess nonvanishing solutions if and only if (\ref{37}) are
no longer closed, and hence not co-cycles. Substituting (\ref{49}--\ref{50})
into (\ref{47}--\ref{48}), it follows that the actions of the irreducible
Koszul-Tate differentials on the initial generators with total resolution
degree one are given by 
\begin{equation}
\delta _{1}\stackrel{[1,0]}{\cal P}_{1a_{0}}=-G_{a_{0}}-A_{a_{0}}^{\;%
\;a_{1}}\pi _{a_{1}},\;\delta _{2}\stackrel{[1,0]}{\cal P}_{1a_{0}}=0,
\label{52}
\end{equation}
\begin{equation}
\delta _{1}\stackrel{[0,1]}{\cal P}_{2a_{0}}=0,\;\delta _{2}\stackrel{[0,1]}%
{\cal P}_{2a_{0}}=-G_{a_{0}}-A_{a_{0}}^{\;\;a_{1}}\pi _{a_{1}},  \label{53}
\end{equation}
which offer us the concrete form of the searched for irreducible constraints
like 
\begin{equation}
\gamma _{a_{0}}\equiv G_{a_{0}}+A_{a_{0}}^{\;\;a_{1}}\pi _{a_{1}}\approx 0.
\label{54}
\end{equation}
The first-class behaviour of the new constraints results from (\ref{54}),
which, after some obvious computation, produce 
\begin{equation}
\pi _{a_{1}}=\bar{D}_{\;\;a_{1}}^{b_{1}}Z_{\;\;b_{1}}^{b_{0}}\gamma
_{b_{0}},\;G_{a_{0}}=\left( \delta _{\;\;a_{0}}^{b_{0}}-Z_{\;\;b_{1}}^{b_{0}}%
\bar{D}_{\;\;a_{1}}^{b_{1}}A_{a_{0}}^{\;\;a_{1}}\right) \gamma _{b_{0}},
\label{54a}
\end{equation}
which consequently yield 
\begin{equation}
\left[ \gamma _{a_{0}},\gamma _{b_{0}}\right] =\bar{C}_{\;%
\;a_{0}b_{0}}^{c_{0}}\gamma _{c_{0}},  \label{54b}
\end{equation}
for some $\bar{C}_{\;\;a_{0}b_{0}}^{c_{0}}$.

In conclusion, we passed from the initial reducible Koszul-Tate bicomplex
associated with the first-stage reducible first-class constraints (\ref{1})
to an irreducible one, based on the generators 
\begin{equation}
\left( \stackrel{[0,0]}{z}^{A},\stackrel{[0,0]}{z}^{A_{1}}\right) ,\;\left( 
\stackrel{[1,0]}{\cal P}_{1a_{0}},\stackrel{[0,1]}{\cal P}_{2a_{0}}\right)
,\;\left( \stackrel{[1,1]}{\lambda }_{a_{0}}\right) ,  \label{55}
\end{equation}
and the definitions 
\begin{equation}
\delta _{1}z^{A}=0,\;\delta _{2}z^{A}=0,\;\delta _{1}z^{A_{1}}=0,\;\delta
_{2}z^{A_{1}}=0,  \label{56}
\end{equation}
\begin{equation}
\delta _{1}\stackrel{[1,0]}{\cal P}_{1a_{0}}=-\gamma _{a_{0}},\;\delta _{2}%
\stackrel{[1,0]}{\cal P}_{1a_{0}}=0,  \label{57}
\end{equation}
\begin{equation}
\delta _{1}\stackrel{[0,1]}{\cal P}_{2a_{0}}=0,\;\delta _{2}\stackrel{[0,1]}%
{\cal P}_{2a_{0}}=-\gamma _{a_{0}},  \label{58}
\end{equation}
\begin{equation}
\delta _{1}\stackrel{[1,1]}{\lambda }_{a_{0}}=-\stackrel{[0,1]}{\cal P}%
_{2a_{0}},\;\delta _{2}\stackrel{[1,1]}{\lambda }_{a_{0}}=\stackrel{[1,0]}%
{\cal P}_{1a_{0}},  \label{59}
\end{equation}
where the irreducible first-class constraint functions $\gamma _{a_{0}}$ are
expressed by (\ref{54}). It can be simply checked that the relations (\ref
{56}--\ref{59}) define a correct Koszul-Tate bicomplex, that satisfies the
basic requirements (\ref{14a}--\ref{14c}), with $\delta _{Ra}$ and $%
C^{\infty }\left( \Sigma \right) $ replaced by $\delta _{a}$, respectively, $%
C^{\infty }\left( \Sigma ^{\prime }\right) $, where $\Sigma ^{\prime
}:\gamma _{a_{0}}\approx 0$.

At this moment it is important to make two essential remarks. Firstly, the
number of physical degrees of freedom is kept unchanged when passing to the
irreducible setting. This is because in the reducible case there are $N$
canonical variables and $M_{0}-M_{1}$ independent first-class constraints,
and hence $\frac{N}{2}-M_{0}+M_{1}$ physical degrees of freedom, while in
the irreducible situation there are $N+2M_{1}$ canonical variables and $M_{0}
$ independent first-class constraints, and therefore as many physical
degrees of freedom as in the reducible version. Secondly, from (\ref{49}--%
\ref{50}) it results (due to the invertibility of $D_{\;\;a_{1}}^{b_{1}}$)
that the momenta $\pi _{a_{1}}$ are both $\delta _{1}$- and $\delta _{2}$%
-exact. These observations represent two main conditions to also be imposed
in connection with higher-order reducible Hamiltonian systems. Actually, we
will see that for higher-order reducible theories it will be necessary to
further add some bosonic canonical variables with the resolution bidegrees $%
\left( 0,0\right) $ or $\left( 1,1\right) $, and some fermionic antighosts
of total resolution degree one. The former condition is directly correlated
to the number of new canonical pairs and first-class constraints needed
within the irreducible framework, while the latter implements the existence
of a proper redefinition of the antighosts with total resolution degree one
that renders trivial the co-cycles from the reducible approach.

\subsection{Second-stage reducible theories}

In this situation we start from an on-shell second-stage reducible set of
first-class constraints ($L=2$), pictured by the formulae (\ref{1}--\ref{4}%
). The reducible Koszul-Tate bicomplex is now generated by (\ref{15}--\ref
{15a}) and (\ref{25}). Obviously, the co-cycles of $\left( \delta
_{Ra}\right) _{a=1,2}$ at total resolution degree one exclusively due to the
original reducibility relations (\ref{3}--\ref{4}) are those appearing in
the first-stage case, namely, (\ref{37}). We are going to maintain the
definitions (\ref{56}--\ref{59}), with $\gamma _{0}$ given precisely by (\ref
{54}). Still, we have to surpass two obstructions that prevent us from
employing the analysis realized in the case $L=1$. Firstly, the matrix $%
D_{\;\;a_{1}}^{b_{1}}=Z_{\;\;a_{1}}^{a_{0}}A_{a_{0}}^{\;\;b_{1}}$ is no
longer invertible, so formula (\ref{42}) is meaningless, such that the
transformations (\ref{38}) are inadequate. This is because $%
D_{\;\;a_{1}}^{b_{1}}$ displays on-shell null vectors 
\begin{equation}
D_{\;\;a_{1}}^{b_{1}}Z_{\;\;a_{2}}^{a_{1}}=A_{a_{0}}^{\;\;b_{1}}Z_{\;%
\;a_{1}}^{a_{0}}Z_{\;\;a_{2}}^{a_{1}}=A_{a_{0}}^{\;%
\;b_{1}}C_{a_{2}}^{a_{0}b_{0}}G_{b_{0}}\approx 0,  \label{60}
\end{equation}
hence now we have that $rank\left( D_{\;\;a_{1}}^{b_{1}}\right) =M_{1}-M_{2}$
. Under these circumstances, we will see that there is, however, possible to
perform an appropriate redefinition of the antighosts $\stackrel{[1,0]}{\cal %
P}_{1a_{0}}$ and $\stackrel{[0,1]}{\cal P}_{2a_{0}}$ that brings the
constraint functions $\gamma _{a_{0}}$ to the form (\ref{54}) and, in the
meantime, restores the triviality of the co-cycles of the type (\ref{37}).
Secondly, we observe that the irreducible constraint functions (\ref{54})
cannot ensure a number of physical degrees of freedom in the irreducible
framework equal with the initial one. Rather, they should be supplemented
with $M_{2}$ new constraints $\gamma _{a_{2}}\approx 0$ such that the entire
constraint set is irreducible and first-class. Accordingly, we have to
enrich the resulting irreducible Koszul-Tate bicomplex with the new
antighosts $\stackrel{[1,0]}{\cal P}_{1a_{2}}$, $\stackrel{[0,1]}{\cal P}%
_{2a_{2}}$ and $\stackrel{[1,1]}{\lambda }_{a_{2}}$ and set the
corresponding definitions 
\begin{equation}
\delta _{1}\stackrel{[1,0]}{\cal P}_{1a_{2}}=-\gamma _{a_{2}},\;\delta _{2}%
\stackrel{[1,0]}{\cal P}_{1a_{2}}=0,  \label{61}
\end{equation}
\begin{equation}
\delta _{1}\stackrel{[0,1]}{\cal P}_{2a_{2}}=0,\;\delta _{2}\stackrel{[0,1]}%
{\cal P}_{2a_{2}}=-\gamma _{a_{2}},  \label{62}
\end{equation}
\begin{equation}
\delta _{1}\stackrel{[1,1]}{\lambda }_{a_{2}}=-\stackrel{[0,1]}{\cal P}%
_{2a_{2}},\;\delta _{2}\stackrel{[1,1]}{\lambda }_{a_{2}}=\stackrel{[1,0]}%
{\cal P}_{1a_{2}}.  \label{63}
\end{equation}
The appearance of the antighosts $\stackrel{[1,1]}{\lambda }_{a_{2}}$ is
dictated by the latter and former definitions in (\ref{61}), respectively, (%
\ref{62}), and is strictly demanded by the supplementary trivial redundancy
in the Hamiltonian BRST-anti-BRST formalism due to the doubling of the
constraint functions. In brief, our programme in the case $L=2$ consists in
determining some $\gamma _{a_{2}}$ that restore a correct irreducible
Koszul-Tate bicomplex relying on the definitions (\ref{56}--\ref{59}) and (%
\ref{61}--\ref{63}), and in further finding a transformation of the
antighosts $\stackrel{[1,0]}{\cal P}_{1a_{0}}$, $\stackrel{[0,1]}{\cal P}%
_{2a_{0}}$ (eventually involving the additional antighosts) that leads to
some trivial co-cycles of the type (\ref{37}) and, at the same time, is in
agreement with (\ref{57}--\ref{58}).

In order to solve the former problem, we remark that condition (\ref{60})
enables us to represent $D_{\;\;a_{1}}^{b_{1}}$ as 
\begin{equation}
D_{\;\;a_{1}}^{b_{1}}=\delta _{\;\;a_{1}}^{b_{1}}-Z_{\;\;b_{2}}^{b_{1}}\bar{D%
}_{\;\;a_{2}}^{b_{2}}A_{a_{1}}^{\;\;a_{2}}+A_{a_{0}}^{\;%
\;b_{1}}C_{c_{2}}^{a_{0}b_{0}}\bar{D}_{\;\;b_{2}}^{c_{2}}A_{a_{1}}^{\;%
\;b_{2}}G_{b_{0}},  \label{64}
\end{equation}
where $\bar{D}_{\;\;a_{2}}^{b_{2}}$ is the inverse of $D_{\;%
\;a_{2}}^{b_{2}}=Z_{\;\;a_{2}}^{c_{1}}A_{c_{1}}^{\;\;b_{2}}$ and $%
A_{c_{1}}^{\;\;b_{2}}$ are some functions that may depend at most on $z^{A}$
and are taken to fulfill $rank\left( D_{\;\;b_{2}}^{a_{2}}\right) =M_{2}$.
As we have previously stated, we maintain the definitions (\ref{57}--\ref{58}%
), with $\gamma _{a_{0}}$ as in (\ref{54}). Then, if we apply $%
Z_{\;\;a_{1}}^{a_{0}}$ on these relations, we consequently find 
\begin{equation}
\delta _{1}\left( \stackrel{[1,0]}{\cal P}_{1a_{0}}Z_{\;\;a_{1}}^{a_{0}}%
\right) =-\pi _{b_{1}}D_{\;\;a_{1}}^{b_{1}},  \label{65}
\end{equation}
\begin{equation}
\delta _{2}\left( \stackrel{[0,1]}{\cal P}_{2a_{0}}Z_{\;\;a_{1}}^{a_{0}}%
\right) =-\pi _{b_{1}}D_{\;\;a_{1}}^{b_{1}}.  \label{66}
\end{equation}
Taking into account the representation (\ref{64}), formulae (\ref{65}--\ref
{66}) become 
\begin{equation}
\delta _{1}\left( \stackrel{[1,0]}{\cal P}_{1a_{0}}Z_{\;\;a_{1}}^{a_{0}}%
\right) =-\pi _{a_{1}}+\pi _{b_{1}}Z_{\;\;b_{2}}^{b_{1}}\bar{D}%
_{\;\;a_{2}}^{b_{2}}A_{a_{1}}^{\;\;a_{2}}-\pi
_{b_{1}}A_{a_{0}}^{\;\;b_{1}}C_{c_{2}}^{a_{0}b_{0}}G_{b_{0}}\bar{D}%
_{\;\;b_{2}}^{c_{2}}A_{a_{1}}^{\;\;b_{2}},  \label{67}
\end{equation}
\begin{equation}
\delta _{2}\left( \stackrel{[0,1]}{\cal P}_{2a_{0}}Z_{\;\;a_{1}}^{a_{0}}%
\right) =-\pi _{a_{1}}+\pi _{b_{1}}Z_{\;\;b_{2}}^{b_{1}}\bar{D}%
_{\;\;a_{2}}^{b_{2}}A_{a_{1}}^{\;\;a_{2}}-\pi
_{b_{1}}A_{a_{0}}^{\;\;b_{1}}C_{c_{2}}^{a_{0}b_{0}}G_{b_{0}}\bar{D}%
_{\;\;b_{2}}^{c_{2}}A_{a_{1}}^{\;\;b_{2}},  \label{68}
\end{equation}
which further yield 
\begin{equation}
\delta _{1}\left( \stackrel{[1,0]}{\cal P}_{1a_{0}}\left(
Z_{\;\;a_{1}}^{a_{0}}-C_{c_{2}}^{a_{0}b_{0}}G_{b_{0}}\bar{D}%
_{\;\;b_{2}}^{c_{2}}A_{a_{1}}^{\;\;b_{2}}\right) \right) =-\pi _{a_{1}}+\pi
_{b_{1}}Z_{\;\;b_{2}}^{b_{1}}\bar{D}_{\;\;a_{2}}^{b_{2}}A_{a_{1}}^{\;%
\;a_{2}},  \label{69}
\end{equation}
\begin{equation}
\delta _{2}\left( \stackrel{[0,1]}{\cal P}_{2a_{0}}\left(
Z_{\;\;a_{1}}^{a_{0}}-C_{c_{2}}^{a_{0}b_{0}}G_{b_{0}}\bar{D}%
_{\;\;b_{2}}^{c_{2}}A_{a_{1}}^{\;\;b_{2}}\right) \right) =-\pi _{a_{1}}+\pi
_{b_{1}}Z_{\;\;b_{2}}^{b_{1}}\bar{D}_{\;\;a_{2}}^{b_{2}}A_{a_{1}}^{\;%
\;a_{2}}.  \label{70}
\end{equation}
The latter two relations are implied by (\ref{67}--\ref{68}) through the
antisymmetry of $C_{c_{2}}^{a_{0}b_{0}}$ (from which it results that $\pi
_{b_{1}}A_{a_{0}}^{\;\;b_{1}}C_{c_{2}}^{a_{0}b_{0}}G_{b_{0}}=\gamma
_{a_{0}}C_{c_{2}}^{a_{0}b_{0}}G_{b_{0}}$) combined with formulae (\ref{56}--%
\ref{58}). The presence of the second term in the right-hand sides of (\ref
{69}--\ref{70}) shows that the momenta $\pi _{a_{1}}$ are neither $\delta
_{1}$-, nor $\delta _{2}$-exact. The most natural choice to surpass this
difficulty is to take the searched for functions $\gamma _{a_{2}}$ as 
\begin{equation}
\gamma _{a_{2}}\equiv \pi _{b_{1}}Z_{\;\;a_{2}}^{b_{1}},  \label{71}
\end{equation}
such that with the help of (\ref{61}--\ref{62}) we arrive at 
\begin{equation}
\delta _{1}\stackrel{[1,0]}{\cal P}_{1a_{2}}=-\pi
_{a_{1}}Z_{\;\;a_{2}}^{a_{1}},\;\delta _{2}\stackrel{[0,1]}{\cal P}%
_{2a_{2}}=-\pi _{a_{1}}Z_{\;\;a_{2}}^{a_{1}}.  \label{72}
\end{equation}
Accordingly, from (\ref{69}--\ref{70}) and (\ref{72}) we recover the $\left(
\delta _{a}\right) _{a=1,2}$-exactness of these momenta 
\begin{equation}
\pi _{a_{1}}=\delta _{1}\left( \stackrel{[1,0]}{\cal P}_{1a_{0}}\left(
-Z_{\;\;a_{1}}^{a_{0}}+C_{c_{2}}^{a_{0}b_{0}}G_{b_{0}}\bar{D}%
_{\;\;b_{2}}^{c_{2}}A_{a_{1}}^{\;\;b_{2}}\right) -\stackrel{[1,0]}{\cal P}%
_{1b_{2}}\bar{D}_{\;\;a_{2}}^{b_{2}}A_{a_{1}}^{\;\;a_{2}}\right) ,
\label{73}
\end{equation}
\begin{equation}
\pi _{a_{1}}=\delta _{2}\left( \stackrel{[0,1]}{\cal P}_{2a_{0}}\left(
-Z_{\;\;a_{1}}^{a_{0}}+C_{c_{2}}^{a_{0}b_{0}}G_{b_{0}}\bar{D}%
_{\;\;b_{2}}^{c_{2}}A_{a_{1}}^{\;\;b_{2}}\right) -\stackrel{[0,1]}{\cal P}%
_{2b_{2}}\bar{D}_{\;\;a_{2}}^{b_{2}}A_{a_{1}}^{\;\;a_{2}}\right) .
\label{74}
\end{equation}
In the meantime, as the functions $Z_{\;\;a_{2}}^{a_{1}}$ have no null
vectors (the original set of constraints is by assumption second-stage
reducible, so $Z_{\;\;a_{2}}^{a_{1}}$ are supposed to be independent),
relations (\ref{72}) provoke no nontrivial co-cycles. This solves the former
problem set in the above in an appropriate manner.

Related to the latter problem (the existence of a transformation of the
antighosts at total resolution degree one outputting some trivial co-cycles
of the type (\ref{37}) and being in accordance with (\ref{57}--\ref{58})),
we observe that replacing the relations (\ref{73}--\ref{74}) in the
definitions (\ref{57}--\ref{58}), with $\gamma _{a_{0}}$ given by (\ref{54}%
), it follows that 
\begin{equation}
\delta _{1}\stackrel{[1,0]}{{\cal P}^{\prime \prime }}_{1a_{0}}=-G_{a_{0}},%
\;\delta _{2}\stackrel{[0,1]}{{\cal P}^{\prime \prime }}_{2a_{0}}=-G_{a_{0}},
\label{75}
\end{equation}
where 
\begin{eqnarray}
&&\stackrel{[1,0]}{{\cal P}^{\prime \prime }}_{1a_{0}}=\stackrel{[1,0]}{\cal %
P}_{1a_{0}}-\stackrel{[1,0]}{\cal P}_{1b_{0}}Z_{\;%
\;b_{1}}^{b_{0}}A_{a_{0}}^{\;\;b_{1}}+  \nonumber  \label{76} \\
&&\stackrel{[1,0]}{\cal P}_{1b_{0}}C_{c_{2}}^{b_{0}c_{0}}G_{c_{0}}\bar{D}%
_{\;\;b_{2}}^{c_{2}}A_{a_{1}}^{\;\;b_{2}}A_{a_{0}}^{\;\;a_{1}}-\stackrel{%
[1,0]}{\cal P}_{1b_{2}}\bar{D}_{\;\;a_{2}}^{b_{2}}A_{a_{1}}^{\;%
\;a_{2}}A_{a_{0}}^{\;\;a_{1}},
\end{eqnarray}
\begin{eqnarray}
&&\stackrel{[0,1]}{{\cal P}^{\prime \prime }}_{2a_{0}}=\stackrel{[0,1]}{\cal %
P}_{2a_{0}}-\stackrel{[0,1]}{\cal P}_{2b_{0}}Z_{\;%
\;b_{1}}^{b_{0}}A_{a_{0}}^{\;\;b_{1}}+  \nonumber  \label{77} \\
&&\stackrel{[0,1]}{\cal P}_{2b_{0}}C_{c_{2}}^{b_{0}c_{0}}G_{c_{0}}\bar{D}%
_{\;\;b_{2}}^{c_{2}}A_{a_{1}}^{\;\;b_{2}}A_{a_{0}}^{\;\;a_{1}}-\stackrel{%
[0,1]}{\cal P}_{2b_{2}}\bar{D}_{\;\;a_{2}}^{b_{2}}A_{a_{1}}^{\;%
\;a_{2}}A_{a_{0}}^{\;\;a_{1}}.
\end{eqnarray}
After some computation, we gain the triviality of the new co-cycles 
\begin{equation}
\mu _{1a_{1}}^{\prime \prime }=\stackrel{[1,0]}{{\cal P}^{\prime \prime }}%
_{1a_{0}}Z_{\;\;a_{1}}^{a_{0}},\;\mu _{2a_{1}}^{\prime \prime }=\stackrel{%
[0,1]}{{\cal P}^{\prime \prime }}_{2a_{0}}Z_{\;\;a_{1}}^{a_{0}},  \label{78}
\end{equation}
at total resolution degree one that result from (\ref{75}), namely, 
\begin{equation}
\mu _{1a_{1}}^{\prime \prime }=\delta _{1}\left( \frac{1}{2}\stackrel{[1,0]}{%
{\cal P}^{\prime \prime }}_{1b_{0}}\stackrel{[1,0]}{{\cal P}^{\prime \prime }%
}_{1a_{0}}C_{c_{2}}^{a_{0}b_{0}}\bar{D}_{\;\;b_{2}}^{c_{2}}A_{a_{1}}^{\;%
\;b_{2}}\right) ,  \label{79}
\end{equation}
\begin{equation}
\mu _{2a_{1}}^{\prime \prime }=\delta _{2}\left( \frac{1}{2}\stackrel{[0,1]}{%
{\cal P}^{\prime \prime }}_{2b_{0}}\stackrel{[0,1]}{{\cal P}^{\prime \prime }%
}_{2a_{0}}C_{c_{2}}^{a_{0}b_{0}}\bar{D}_{\;\;b_{2}}^{c_{2}}A_{a_{1}}^{\;%
\;b_{2}}\right) .  \label{80}
\end{equation}
Formulae (\ref{76}--\ref{77}) help us to determine the concrete form of the
redefinition under discussion 
\begin{equation}
\left( \stackrel{\lbrack 1,0]}{\cal P}_{1a_{0}},\stackrel{[1,0]}{\cal P}%
_{1a_{2}},\stackrel{[0,1]}{\cal P}_{2a_{0}},\stackrel{[0,1]}{\cal P}%
_{2a_{2}}\right) \rightarrow \left( \stackrel{[1,0]}{{\cal P}^{\prime \prime
}}_{1a_{0}},\stackrel{[1,0]}{\cal P}_{1a_{2}},\stackrel{[0,1]}{{\cal P}%
^{\prime \prime }}_{2a_{0}},\stackrel{[0,1]}{\cal P}_{2a_{2}}\right) ,
\label{81}
\end{equation}
where 
\begin{equation}
\stackrel{\lbrack 1,0]}{{\cal P}^{\prime \prime }}_{1a_{0}}=R_{\;%
\;a_{0}}^{b_{0}}\stackrel{[1,0]}{\cal P}_{1b_{0}}+Q_{\;\;a_{0}}^{b_{2}}%
\stackrel{[1,0]}{\cal P}_{1b_{2}},  \label{82}
\end{equation}
\begin{equation}
\stackrel{\lbrack 0,1]}{{\cal P}^{\prime \prime }}_{2a_{0}}=R_{\;%
\;a_{0}}^{b_{0}}\stackrel{[0,1]}{\cal P}_{2b_{0}}+Q_{\;\;a_{0}}^{b_{2}}%
\stackrel{[0,1]}{\cal P}_{2b_{2}},  \label{83}
\end{equation}
\begin{equation}
R_{\;\;a_{0}}^{b_{0}}=\delta
_{\;\;a_{0}}^{b_{0}}-Z_{\;\;b_{1}}^{b_{0}}A_{a_{0}}^{\;%
\;b_{1}}+C_{c_{2}}^{b_{0}c_{0}}G_{c_{0}}\bar{D}_{\;%
\;b_{2}}^{c_{2}}A_{a_{1}}^{\;\;b_{2}}A_{a_{0}}^{\;\;a_{1}},Q_{\;%
\;a_{0}}^{b_{2}}=-\bar{D}_{\;\;a_{2}}^{b_{2}}A_{a_{1}}^{\;%
\;a_{2}}A_{a_{0}}^{\;\;a_{1}}.  \label{84}
\end{equation}
As no nontrivial co-cycles connected with the initial reducibility appear at
total resolution degree one, it follows that there will also be no
nontrivial co-cycles correlated to this type of reducibility at higher
resolution degrees. Consequently, the constraints underlying the new
Koszul-Tate bicomplex 
\begin{equation}
\gamma _{a_{0}}\equiv G_{a_{0}}+A_{a_{0}}^{\;\;a_{1}}\pi _{a_{1}}\approx
0,\;\gamma _{a_{2}}\equiv Z_{\;\;a_{2}}^{a_{1}}\pi _{a_{1}}\approx 0,
\label{85}
\end{equation}
are truly irreducible. Moreover, they are also first-class as (\ref{85})
leads, after some computation, to 
\begin{equation}
\pi _{a_{1}}=\gamma _{a_{0}}\left(
Z_{\;\;a_{1}}^{a_{0}}-C_{c_{2}}^{a_{0}b_{0}}G_{b_{0}}\bar{D}%
_{\;\;b_{2}}^{c_{2}}A_{a_{1}}^{\;\;b_{2}}\right) +\gamma _{b_{2}}\bar{D}%
_{\;\;a_{2}}^{b_{2}}A_{a_{1}}^{\;\;a_{2}},  \label{86}
\end{equation}
\begin{eqnarray}
&&G_{a_{0}}=\gamma _{b_{0}}\left( \delta
_{\;\;a_{0}}^{b_{0}}-Z_{\;\;b_{1}}^{b_{0}}A_{a_{0}}^{\;%
\;b_{1}}+C_{c_{2}}^{b_{0}c_{0}}G_{c_{0}}\bar{D}_{\;%
\;b_{2}}^{c_{2}}A_{b_{1}}^{\;\;b_{2}}A_{a_{0}}^{\;\;b_{1}}\right) - 
\nonumber  \label{87} \\
&&\gamma _{b_{2}}\bar{D}_{\;\;a_{2}}^{b_{2}}A_{a_{1}}^{\;\;a_{2}}A_{a_{0}}^{%
\;\;a_{1}}.
\end{eqnarray}
Evaluating the Poisson brackets among the constraint functions $\left(
\gamma _{a_{0}},\gamma _{a_{2}}\right) $ with the help of (\ref{86}--\ref{87}%
), we find that they vanish weakly on the surface $\gamma _{a_{0}}\approx 0$%
, $\gamma _{a_{2}}\approx 0$, hence the constraints (\ref{85}) are
first-class.

Then, we can conclude that in the case $L=2$ we succeeded again in switching
from an original reducible Koszul-Tate bicomplex associated with the
second-stage reducible first-class constraints (\ref{1}) to an irreducible
one, based on the generators 
\begin{equation}
\left( \stackrel{\lbrack 0,0]}{z}^{A},\stackrel{[0,0]}{z}^{A_{1}}\right)
,\;\left( \stackrel{[1,0]}{\cal P}_{1a_{0}},\stackrel{[1,0]}{\cal P}%
_{1a_{2}},\stackrel{[0,1]}{\cal P}_{2a_{0}},\stackrel{[0,1]}{\cal P}%
_{2a_{2}}\right) ,\;\left( \stackrel{[1,1]}{\lambda }_{a_{0}},\stackrel{[1,1]%
}{\lambda }_{a_{2}}\right) ,  \label{88}
\end{equation}
and defined by the relations (\ref{56}--\ref{59}) and (\ref{61}--\ref{63}),
where the irreducible first-class constraint functions $\gamma _{a_{0}}$, $%
\gamma _{a_{2}}$ are given in (\ref{85}). It can be verified directly that
the above-mentioned relations define a correct irreducible Koszul-Tate
bicomplex, which agrees with the basic conditions (\ref{14a}--\ref{14c}),
with $\delta _{Ra}$ and $C^{\infty }\left( \Sigma \right) $ replaced by $%
\delta _{a}$, respectively, $C^{\infty }\left( \Sigma ^{\prime \prime
}\right) $, where $\Sigma ^{\prime \prime }:\gamma _{a_{0}}\approx 0$, $%
\gamma _{a_{2}}\approx 0$. We cannot stress enough that it is precisely the
requirement on the $\delta _{a}$-exactness of the new momenta $\pi _{a_{1}}$
which allows us to deduce an appropriate transformation of the antighosts at
total resolution degree one (see (\ref{81}--\ref{84})) that enforces the
irreducibility.

\subsection{Generalization: $L$-stage reducible theories}

Now, after having analysed in detail the construction of an irreducible
Koszul-Tate bicomplex starting with an original first- or second-stage
reducible set of first-class constraints, we are able to generalize our
irreducible treatment to some initial on-shell $L$-stage reducible
first-class constraints, described by the relations (\ref{1}--\ref{5}).
Going along the same line as before, we enlarge the phase-space with the
bosonic canonical pairs $\stackrel{[0,0]}{z}^{A_{2k+1}}=\left(
y^{a_{2k+1}},\pi _{a_{2k+1}}\right) _{k=0,\cdots ,\Lambda }$ and construct
an irreducible Koszul-Tate bicomplex based on the generators 
\begin{equation}
\left( \stackrel{\lbrack 0,0]}{z}^{A},\left( \stackrel{[0,0]}{z}%
^{A_{2k+1}}\right) _{k=0,\cdots ,\Lambda }\right) ,\left( \stackrel{[1,0]}%
{\cal P}_{1a_{2k}},\stackrel{[0,1]}{\cal P}_{2a_{2k}},\stackrel{[1,1]}{%
\lambda }_{a_{2k}}\right) _{k=0,\cdots ,\Gamma },  \label{89}
\end{equation}
and defined by the relations 
\begin{equation}
\delta _{a}z^{A}=0,\;\delta _{a}z^{A_{2k+1}}=0,\;k=0,\cdots ,\Lambda
,\;a=1,2,  \label{90}
\end{equation}
\begin{equation}
\delta _{1}\stackrel{[1,0]}{\cal P}_{1a_{2k}}=-\gamma _{a_{2k}},\;\delta _{2}%
\stackrel{[1,0]}{\cal P}_{1a_{2k}}=0,\;k=0,\cdots ,\Gamma ,  \label{91}
\end{equation}
\begin{equation}
\delta _{1}\stackrel{[0,1]}{\cal P}_{2a_{2k}}=0,\;\delta _{2}\stackrel{[0,1]}%
{\cal P}_{2a_{2k}}=-\gamma _{a_{2k}},\;k=0,\cdots ,\Gamma ,  \label{92}
\end{equation}
\begin{equation}
\delta _{1}\stackrel{[1,1]}{\lambda }_{a_{2k}}=-\stackrel{[0,1]}{\cal P}%
_{2a_{2k}},\;\delta _{2}\stackrel{[1,1]}{\lambda }_{a_{2k}}=\stackrel{[1,0]}%
{\cal P}_{1a_{2k}},\;k=0,\cdots ,\Gamma ,  \label{93}
\end{equation}
which shows the irreducible constraints 
\begin{equation}
\gamma _{a_{0}}\equiv G_{a_{0}}+A_{a_{0}}^{\;\;a_{1}}\pi _{a_{1}}\approx 0,
\label{94}
\end{equation}
\begin{equation}
\gamma _{a_{2k}}\equiv Z_{\;\;a_{2k}}^{a_{2k-1}}\pi
_{a_{2k-1}}+A_{a_{2k}}^{\;\;a_{2k+1}}\pi _{a_{2k+1}}\approx 0,\;k=1,\cdots
,\Gamma .  \label{95}
\end{equation}
The irreducibility is guaranteed by the presence of the functions $%
A_{a_{k}}^{\;\;a_{k+1}}$ in (\ref{94}--\ref{95}), which may involve at most
the original variables $z^{A}$ and are chosen to satisfy the conditions 
\begin{equation}
rank\left( D_{\;\;b_{k}}^{a_{k}}\right) \approx \sum\limits_{i=k}^{L}\left(
-\right) ^{k+i}M_{i},\;k=1,\cdots ,L-1,  \label{96}
\end{equation}
\begin{equation}
rank\left( D_{\;\;b_{L}}^{a_{L}}\right) =M_{L},  \label{97}
\end{equation}
where $D_{\;\;b_{k}}^{a_{k}}=Z_{\;\;b_{k}}^{a_{k-1}}A_{a_{k-1}}^{\;\;a_{k}}$%
. Acting like in the cases $L=1,2$, after some computation we find the
relations 
\begin{equation}
\pi _{a_{2k+1}}=\nu _{a_{2k+1}}^{a_{2j}}\gamma _{a_{2j}},\;k=0,\cdots
,\Lambda ,  \label{99}
\end{equation}
\begin{equation}
G_{a_{0}}=\nu _{a_{0}}^{a_{2j}}\gamma _{a_{2j}},  \label{100}
\end{equation}
for some functions $\nu _{a_{2k+1}}^{a_{2j}}$ and $\nu _{a_{0}}^{a_{2j}}$.
Computing the Poisson brackets among the constraint functions in (\ref{94}--%
\ref{95}), we find that they vanish weakly on the surface (\ref{94}--\ref{95}%
), hence they form a first-class set. The antighosts of the type $\stackrel{%
[1,0]}{\cal P}_{1a_{2k}}$ or $\stackrel{[0,1]}{\cal P}_{2a_{2k}}$ are
fermionic, those denoted by $\stackrel{[1,1]}{\lambda }_{a_{2k}}$ are
bosonic, while the notations $\Lambda $ and $\Gamma $ signify 
\begin{equation}
\Lambda =\left\{ 
\begin{array}{c}
\frac{L-1}{2},\;{\rm for}\;L\;{\rm odd}, \\ 
\frac{L}{2}-1,\;{\rm for}\;L\;{\rm even},
\end{array}
\right. \;\;\Gamma =\left\{ 
\begin{array}{c}
\frac{L-1}{2},\;{\rm for}\;L\;{\rm odd}, \\ 
\frac{L}{2},\;{\rm for}\;L\;{\rm even}.
\end{array}
\right.   \label{98}
\end{equation}
In order to avoid confusion, we use the conventions $f^{a_{k}}=0$ if $k<0$
or $k>L$.

With these elements at hand, it is obviously that the Koszul-Tate operators
satisfy the fundamental requirements (\ref{14a}--\ref{14c}), with $\delta
_{Ra}$ and $C^{\infty }\left( \Sigma \right) $ replaced by $\delta _{a}$,
respectively, $C^{\infty }\left( \bar{\Sigma}\right) $, where 
\begin{equation}
\bar{\Sigma}:\gamma _{a_{2k}}\approx 0,\;k=0,\cdots ,\Gamma .  \label{irsurf}
\end{equation}
This ends the general construction of an irreducible Koszul-Tate bicomplex
associated with the original on-shell $L$-stage reducible Hamiltonian system.

\section{Irreducible Hamiltonian BRST-anti-BRST symmetry}

\subsection{Construction of the irreducible Hamiltonian BRST-anti-BRST
symmetry}

Once we have accomplished the construction of the Koszul-Tate bicomplex
based on the irreducible first-class constraints (\ref{94}--\ref{95}), it is
natural to derive the irreducible Hamiltonian BRST-anti-BRST symmetry
associated with this constraint set. The BRST and anti-BRST operators $%
\left( s_{a}\right) _{a=1,2}$ corresponding to this irreducible first-class
set should anticommute 
\begin{equation}
s_{a}s_{b}+s_{b}s_{a}=0,\;a,b=1,2,  \label{101}
\end{equation}
each of the two differentials splitting as 
\begin{equation}
s_{a}=\delta _{a}+D_{a}+\cdots ,\;a=1,2,  \label{102}
\end{equation}
where $\left( \delta _{a}\right) _{a=1,2}$ generate the irreducible
Koszul-Tate bicomplex, $\left( D_{a}\right) _{a=1,2}$ define the irreducible
exterior longitudinal bicomplex, and the other terms (if necessary),
generically denoted by ``$\cdots $'', implement the anticommutation
relations (\ref{101}). The construction of the irreducible Koszul-Tate
bicomplex has been elucidated in the previous subsections. With regard to
the irreducible exterior longitudinal bicomplex, we remark that its
generators are given by 
\begin{equation}
\left( \stackrel{\{0,0\}}{z}^{A},\left( \stackrel{\{0,0\}}{z}%
^{A_{2k+1}}\right) _{k=0,\cdots ,\Lambda }\right) ,\;\left( \stackrel{\{1,0\}%
}{\eta }_{1}^{a_{2k}},\stackrel{\{0,1\}}{\eta }_{2}^{a_{2k}},\stackrel{%
\{1,1\}}{Q}^{a_{2k}}\right) _{k=0,\cdots ,\Gamma },  \label{103}
\end{equation}
where $\stackrel{\{1,0\}}{\eta }_{1}^{a_{2k}}$, $\stackrel{\{0,1\}}{\eta }%
_{2}^{a_{2k}}$ are fermionic, $\stackrel{\{1,1\}}{Q}^{a_{2k}}$ are bosonic,
and $\stackrel{\{m,n\}}{F}$ represents an element of form bidegree $%
biform\left( F\right) =(m,n)$. For notational simplicity, we redenote the
first-class constraints (\ref{94}--\ref{95}) by $\gamma _{\bar{\Delta}%
}\equiv \left( \gamma _{a_{2k}}\right) _{k=0,\cdots ,\Gamma }\approx 0$,
such that we can equivalently rewrite the ghost spectrum in (\ref{103}) like 
$\stackrel{\{1,0\}}{\eta }_{1}^{\bar{\Delta}}$, $\stackrel{\{0,1\}}{\eta }%
_{2}^{\bar{\Delta}}$, $\stackrel{\{1,1\}}{Q}^{\bar{\Delta}}$. Due on the one
hand to the first-class character of the constraints (\ref{94}--\ref{95}) 
\begin{equation}
\left[ \gamma _{\bar{\Delta}},\gamma _{\bar{\Delta}^{\prime }}\right] =C_{%
\bar{\Delta}\bar{\Delta}^{\prime }}^{\bar{\Delta}^{\prime \prime }}\gamma _{%
\bar{\Delta}^{\prime \prime }},  \label{104}
\end{equation}
where $C_{\bar{\Delta}\bar{\Delta}^{\prime }}^{\bar{\Delta}^{\prime \prime }}
$ denote the first-order structure functions, and on the other hand, to
their irreducibility, it follows that the actions of $\left( D_{a}\right)
_{a=1,2}$ on the generators from the exterior longitudinal bicomplex are
completely defined through the relations 
\begin{equation}
D_{1}\stackrel{\{0,0\}}{F}=\left[ \stackrel{\{0,0\}}{F},\gamma _{\bar{\Delta}%
}\right] \stackrel{\{1,0\}}{\eta }_{1}^{\bar{\Delta}},\;D_{2}\stackrel{%
\{0,0\}}{F}=\left[ \stackrel{\{0,0\}}{F},\gamma _{\bar{\Delta}}\right] 
\stackrel{\{0,1\}}{\eta }_{2}^{\bar{\Delta}},  \label{105}
\end{equation}
\begin{equation}
D_{1}\stackrel{\{1,0\}}{\eta }_{1}^{\bar{\Delta}}=\frac{1}{2}C_{\bar{\Delta}%
^{\prime }\bar{\Delta}^{\prime \prime }}^{\bar{\Delta}}\stackrel{\{1,0\}}{%
\eta }_{1}^{\bar{\Delta}^{\prime }}\stackrel{\{1,0\}}{\eta }_{1}^{\bar{\Delta%
}^{\prime \prime }},  \label{106}
\end{equation}
\begin{equation}
D_{2}\stackrel{\{1,0\}}{\eta }_{1}^{\bar{\Delta}}=\stackrel{\{1,1\}}{Q}^{%
\bar{\Delta}}+\frac{1}{2}C_{\bar{\Delta}^{\prime }\bar{\Delta}^{\prime
\prime }}^{\bar{\Delta}}\stackrel{\{1,0\}}{\eta }_{1}^{\bar{\Delta}^{\prime
}}\stackrel{\{0,1\}}{\eta }_{2}^{\bar{\Delta}^{\prime \prime }},
\label{106a}
\end{equation}
\begin{equation}
D_{1}\stackrel{\{0,1\}}{\eta }_{2}^{\bar{\Delta}}=-\stackrel{\{1,1\}}{Q}^{%
\bar{\Delta}}+\frac{1}{2}C_{\bar{\Delta}^{\prime }\bar{\Delta}^{\prime
\prime }}^{\bar{\Delta}}\stackrel{\{0,1\}}{\eta }_{2}^{\bar{\Delta}^{\prime
}}\stackrel{\{1,0\}}{\eta }_{1}^{\bar{\Delta}^{\prime \prime }},  \label{107}
\end{equation}
\begin{equation}
D_{2}\stackrel{\{0,1\}}{\eta }_{2}^{\bar{\Delta}}=\frac{1}{2}C_{\bar{\Delta}%
^{\prime }\bar{\Delta}^{\prime \prime }}^{\bar{\Delta}}\stackrel{\{0,1\}}{%
\eta }_{2}^{\bar{\Delta}^{\prime }}\stackrel{\{0,1\}}{\eta }_{2}^{\bar{\Delta%
}^{\prime \prime }},  \label{107a}
\end{equation}
\begin{equation}
D_{1}\stackrel{\{1,1\}}{Q}^{\bar{\Delta}}=\frac{1}{2}C_{\bar{\Delta}^{\prime
}\bar{\Delta}^{\prime \prime }}^{\bar{\Delta}}\stackrel{\{1,1\}}{Q}^{\bar{%
\Delta}^{\prime }}\stackrel{\{1,0\}}{\eta }_{1}^{\bar{\Delta}^{\prime \prime
}}+a^{\bar{\Delta}},\;D_{2}\stackrel{\{1,1\}}{Q}^{\bar{\Delta}}=\frac{1}{2}%
C_{\bar{\Delta}^{\prime }\bar{\Delta}^{\prime \prime }}^{\bar{\Delta}}%
\stackrel{\{1,1\}}{Q}^{\bar{\Delta}^{\prime }}\stackrel{\{0,1\}}{\eta }_{2}^{%
\bar{\Delta}^{\prime \prime }}+b^{\bar{\Delta}},  \label{108}
\end{equation}
where $\stackrel{\{0,0\}}{F}$ can be any function of the variables $z^{A}$, $%
\left( z^{A_{2k+1}}\right) _{k=0,\cdots ,\Lambda }$, and the functions $a^{%
\bar{\Delta}}$ and $b^{\bar{\Delta}}$ read as $a^{\bar{\Delta}}=\frac{1}{8}%
C_{\bar{\Delta}_{1}\bar{\Delta}^{\prime }}^{\bar{\Delta}}C_{\bar{\Delta}_{2}%
\bar{\Delta}_{3}}^{\bar{\Delta}^{\prime }}\stackrel{\{0,1\}}{\eta }_{2}^{%
\bar{\Delta}_{1}}\stackrel{\{1,0\}}{\eta }_{1}^{\bar{\Delta}_{2}}\stackrel{%
\{1,0\}}{\eta }_{1}^{\bar{\Delta}_{3}}$, respectively, $b^{\bar{\Delta}}=-%
\frac{1}{8}C_{\bar{\Delta}_{1}\bar{\Delta}^{\prime }}^{\bar{\Delta}}C_{\bar{%
\Delta}_{2}\bar{\Delta}_{3}}^{\bar{\Delta}^{\prime }}\stackrel{\{1,0\}}{\eta 
}_{1}^{\bar{\Delta}_{1}}\stackrel{\{0,1\}}{\eta }_{2}^{\bar{\Delta}_{2}}%
\stackrel{\{0,1\}}{\eta }_{2}^{\bar{\Delta}_{3}}$. According to the above
definitions, we find that the operators $\left( D_{a}\right) _{a=1,2}$
anticommute weakly 
\begin{equation}
D_{a}D_{b}+D_{b}D_{a}\approx 0,\;a,b=1,2,  \label{109}
\end{equation}
where the weak equality is referring to the first-class surface (\ref{94}--%
\ref{95}). Moreover, we infer that the cohomology spaces of $\left(
D_{a}\right) _{a=1,2}$ at form bidegree $(0,0)$ are given by the algebra of
physical observables corresponding to the irreducible first-class
constraints (\ref{94}--\ref{95}). Then, we infer that the exterior
longitudinal bicomplex associated with the irreducible system meets all the
requirements of the Hamiltonian BRST-anti-BRST method.

If we extend the actions of $\left( \delta _{a}\right) _{a=1,2}$ to the
ghosts through 
\begin{equation}
\delta _{a}\eta _{1}^{\bar{\Delta}}=0,\;\delta _{a}\eta _{2}^{\bar{\Delta}%
}=0,\;\delta _{a}Q^{\bar{\Delta}}=0,  \label{110}
\end{equation}
and set $bires\left( \eta _{1}^{\bar{\Delta}}\right) =\left( 0,0\right) $, $%
bires\left( \eta _{2}^{\bar{\Delta}}\right) =\left( 0,0\right) $, $%
bires\left( Q^{\bar{\Delta}}\right) =\left( 0,0\right) $, we rediscover the
relations 
\begin{equation}
\delta _{a}\delta _{b}+\delta _{b}\delta _{a}=0,\;a,b=1,2,  \label{111}
\end{equation}
\begin{equation}
H_{0,0}\left( \delta _{a}\right) =C^{\infty }\left( \bar{\Sigma}\right)
,\;a=1,2,  \label{112}
\end{equation}
\begin{equation}
H_{j,k}\left( \delta _{a}\right) =0,\;j,k\geq 0,\;j+k\neq 0,\;a=1,2,
\label{113}
\end{equation}
which confirms that the irreducible Koszul-Tate differentials can be
properly prolonged to the entire BRST-anti-BRST generator algebra, i.e. such
that to preserve the biresolution property. Furthermore, the first-class
character of the irreducible constraints (\ref{94}--\ref{95}) ensures that
the irreducible exterior longitudinal derivatives $\left( D_{a}\right)
_{a=1,2}$ (and their accompanying bigrading, $biform$) can also be extended
to the antighosts in such a way as to become differentials modulo $\left(
\delta _{a}\right) _{a=1,2}$. Consequently, we can consistently construct 
\cite{5} the Hamiltonian BRST-anti-BRST symmetry of the irreducible theory,
realized via two anticommuting differentials $\left( s_{a}\right) _{a=1,2}$ 
\begin{equation}
s_{a}s_{b}+s_{b}s_{a}=0,\;a,b=1,2,  \label{114}
\end{equation}
that start like 
\begin{equation}
s_{a}=\delta _{a}+D_{a}+\cdots ,\;a=1,2,  \label{115}
\end{equation}
and are bigraded accordingly with the new ghost bidegree $bingh=\left(
ngh_{1},ngh_{2}\right) $, defined like in (\ref{8}). The generators of the
superalgebra underlying the irreducible Hamiltonian BRST-anti-BRST bicomplex
are precisely (\ref{89}) and the ghosts in (\ref{103}), but bigraded in
terms of $bingh$. Then, the cohomology groups of $\left( s_{a}\right)
_{a=1,2}$ at new ghost bidegree zero, $H^{0,0}\left( s_{a}\right) $, will be
isomorphic to the algebra of physical observables corresponding to the
irreducible system. In the following we will denote by $\stackrel{(m,n)}{F}$
an object with $bingh\left( F\right) =\left( m,n\right) $.

\subsection{Relation with the reducible Hamiltonian BRST-anti-BRST symmetry}

Let us investigate now the relation between the Hamiltonian BRST-anti-BRST
symmetries of the reducible and irreducible systems. In view of this, we
show that the physical observables of the two theories coincide. Let $F$ be
an observable of the irreducible system. Consequently, it satisfies the
equations 
\begin{equation}
\left[ F,\gamma _{a_{2k}}\right] \approx 0,\;k=0,\cdots ,\Gamma ,
\label{x55}
\end{equation}
where the weak equality refers to the surface $\bar{\Sigma}$, given by (\ref
{irsurf}). Using the relations (\ref{99}--\ref{100}), we then find that $F$
also fulfils the equations 
\begin{equation}
\left[ F,G_{a_{0}}\right] =\left[ F,\nu _{a_{0}}^{a_{2j}}\right] \gamma
_{a_{2j}}+\left[ F,\gamma _{a_{2j}}\right] \nu _{a_{0}}^{a_{2j}}\approx 0,
\label{x55a}
\end{equation}
\begin{equation}
\left[ F,\pi _{a_{2k+1}}\right] =\left[ F,\nu _{a_{2k+1}}^{a_{2j}}\right]
\gamma _{a_{2j}}+\left[ F,\gamma _{a_{2j}}\right] \nu
_{a_{2k+1}}^{a_{2j}}\approx 0,\;k=0,\cdots ,\Lambda ,  \label{x55b}
\end{equation}
on this surface. So, every observable of the irreducible theory should
verify the equations (\ref{x55a}--\ref{x55b}) on $\bar{\Sigma}$. Now, we
observe that this surface is equivalent to that described by the equations 
\begin{equation}
G_{a_{0}}\approx 0,\;\pi _{a_{2k+1}}\approx 0,\;k=0,\cdots ,\Lambda .
\label{x55c}
\end{equation}
Indeed, it is clear that if (\ref{x55c}) takes place, then (\ref{94}--\ref
{95}) hold. The converse results from (\ref{99}--\ref{100}), which show that
if (\ref{94}--\ref{95}) hold, then (\ref{x55c}) are also valid. This proves
the equivalence between the first-class surfaces $\bar{\Sigma}$ and (\ref
{x55c}). Consequently, we have that every observable of the irreducible
theory, which we found that verifies equations (\ref{x55a}--\ref{x55b}) on $%
\bar{\Sigma}$, will check these equations also on the surface (\ref{x55c} ).
This means that every observable of the irreducible system is an observable
of the theory based on the first-class constraints (\ref{x55c}). Conversely,
if $F$ represents a physical observable of the system subject to the
constraints (\ref{x55c}), then it should check the equations 
\begin{equation}
\left[ F,G_{a_{0}}\right] \approx 0,\;\left[ F,\pi _{a_{2k+1}}\right]
\approx 0,\;k=0,\cdots ,\Lambda ,  \label{x55d}
\end{equation}
on the surface (\ref{x55c}). Then, on behalf of (\ref{94}--\ref{95}) it
follows that $F$ satisfies the relations 
\begin{equation}
\left[ F,\gamma _{a_{0}}\right] =\left[ F,G_{a_{0}}\right] +\left[
F,A_{a_{0}}^{\;\;a_{1}}\right] \pi _{a_{1}}+\left[ F,\pi _{a_{1}}\right]
A_{a_{0}}^{\;\;a_{1}}\approx 0,  \label{x55e}
\end{equation}
\begin{eqnarray}
&&\left[ F,\gamma _{a_{2k}}\right] =\left[
F,Z_{\;\;a_{2k}}^{a_{2k-1}}\right] \pi _{a_{2k-1}}+\left[ F,\pi
_{a_{2k-1}}\right] Z_{\;\;a_{2k}}^{a_{2k-1}}+  \nonumber  \label{55f} \\
&&\left[ F,A_{a_{2k}}^{\;\;a_{2k+1}}\right] \pi _{a_{2k+1}}+\left[ F,\pi
_{a_{2k+1}}\right] A_{a_{2k}}^{\;\;a_{2k+1}}\approx 0,\;k=1,\cdots ,\Gamma ,
\end{eqnarray}
on the same surface. Recalling once again the equivalence between this
surface and $\bar{\Sigma}$, we find that $F$ will verify the equations (\ref
{x55e}--\ref{55f}) also on $\bar{\Sigma}$, and is therefore an observable of
the irreducible system. From the above discussion we conclude that the
physical observables of the irreducible theory coincide with those
associated with the system subject to the first-class constraints (\ref{x55c}%
). Next, we show that the physical observables of the system possessing the
constraints (\ref{x55c}) and those corresponding to the original reducible
theory coincide. In this light, we remark that the surface (\ref{x55c}) can
be inferred in a trivial manner from (\ref{1}) by adding the canonical pairs 
$\left( y^{a_{2k+1}},\pi _{a_{2k+1}}\right) _{k=0,\cdots ,\Lambda }$ and
requiring that their momenta vanish. Thus, the observables of the original
redundant theory are unaffected by the introduction of the new canonical
pairs. In fact, the difference between an observable $F$ of the system
subject to the constraints (\ref{x55c}) and one of the original theory, $%
\bar{F}$, is of the type $F-\bar{F}=\sum_{k=0}^{\Lambda }f^{a_{2k+1}}\pi
_{a_{2k+1}}$. As any two observables that differ through a combination of
first-class constraint functions can be identified, we find that the
physical observables of the initial theory coincide with those of the system
described by the constraints (\ref{x55c}). So far, we proved that the
observables of the system with the constraints (\ref{x55c}) coincide on the
one hand with those of the irreducible theory, and, on the other hand, with
those of the original reducible one. In conclusion, the physical observables
associated with the irreducible system also coincide with those of the
starting on-shell reducible first-class theory. In turn, this result will
have a strong impact at the level of the BRST-anti-BRST analysis.

In the above we have shown that starting with an arbitrary on-shell
reducible first-class Hamiltonian system displaying the Hamiltonian
BRST-anti-BRST symmetry $\left( s_{Ra}\right) _{a=1,2}$ we can construct a
corresponding irreducible first-class theory, whose BRST-anti-BRST symmetry $%
\left( s_{a}\right) _{a=1,2}$ complies with the basic requirements of the
Hamiltonian BRST-anti-BRST formalism. The previous result on the physical
observables induces that the zeroth order cohomological groups of the
reducible and irreducible BRST and anti-BRST operators are isomorphic 
\begin{equation}
H^{0,0}\left( s_{Ra}\right) \simeq H^{0,0}\left( s_{b}\right) ,\;a,b=1,2.
\label{x56}
\end{equation}
In addition, each symmetry is generated by a couple of anticommuting
differentials 
\begin{equation}
s_{Ra}s_{Rb}+s_{Rb}s_{Ra}=0=s_{a}s_{b}+s_{b}s_{a},\;a,b=1,2.  \label{x57}
\end{equation}
Then, from the point of view of the fundamental equations of the
BRST-anti-BRST formalism, namely, the nilpotency and anticommutativity of
the BRST and anti-BRST operators, respectively, the isomorphism between the
zeroth order cohomological groups of the BRST and anti-BRST differentials
and the algebra of physical observables, it follows that it is permissible
to replace the Hamiltonian BRST-anti-BRST symmetry of the original $L$-stage
reducible system with that of the irreducible theory. Thus, we can
substitute the path integral of the reducible system in the Hamiltonian
BRST-anti-BRST approach by that of the irreducible theory.

However, it would be convenient to infer a covariant path integral with
respect to the irreducible system. The present phase-space coordinates may
not ensure the covariance. For instance, if we analyse the gauge
transformations of the extended action of the irreducible system, we remark
that those corresponding to the Lagrange multipliers of the constraint
functions $\gamma _{a_{0}}$ will not involve the term $-Z_{\;%
\;a_{1}}^{a_{0}}\epsilon ^{a_{1}}$, which is present within the reducible
context with respect to the constraint functions $G_{a_{0}}$. For all known
models, the presence of this term is essential in arriving at some covariant
gauge transformations at the Lagrangian level. For this reason it is
necessary to gain such a term also within the irreducible setting. Moreover,
it is possible that some of the newly added canonical variables lack
covariant Lagrangian gauge transformations. This signalizes that we need to
add more phase-space variables to be constrained in an appropriate manner.
In view of this, we introduce the additional bosonic canonical pairs 
\begin{equation}
\left( y^{(1)a_{2k+1}},\pi _{a_{2k+1}}^{(1)}\right) ,\;\left(
y^{(2)a_{2k+1}},\pi _{a_{2k+1}}^{(2)}\right) ,\;k=0,\cdots ,\Lambda ,
\label{x57aa}
\end{equation}
\begin{equation}
\left( y^{a_{2k}},\pi _{a_{2k}}\right) ,\;k=1,\cdots ,\Gamma ,  \label{x57ab}
\end{equation}
subject to some constraints of the type 
\begin{equation}
\gamma _{a_{2k+1}}^{(1)}\equiv \pi _{a_{2k+1}}-\pi _{a_{2k+1}}^{(1)}\approx
0,\;\gamma _{a_{2k+1}}^{(2)}\equiv \pi _{a_{2k+1}}^{(2)}\approx
0,\;k=0,\cdots ,\Lambda ,  \label{x57a}
\end{equation}
\begin{equation}
\gamma _{a_{2k}}^{(1)}\equiv \pi _{a_{2k}}\approx 0,\;k=1,\cdots ,\Gamma .
\label{x57abc}
\end{equation}
In this manner we do not affect in any way the properties of the irreducible
theory as (\ref{x57a}--\ref{x57abc}) still form together with (\ref{94}--\ref
{95}) an irreducible first-class set. The newly added constraints implies
the introduction of some supplementary ghosts and antighosts and the
extension of the action of the BRST and anti-BRST operators on them in the
usual manner. Then, there exists a consistent Hamiltonian BRST-anti-BRST
symmetry with respect to the new irreducible theory, described by the
constraints (\ref{94}--\ref{95}) and (\ref{x57a}--\ref{x57abc}). Now, if we
choose the first-class Hamiltonian with respect to the above first-class
constraints in an adequate manner, we can, in principle, generate a gauge
algebra that leads to some covariant Lagrangian gauge transformations. >From (%
\ref{99}) it results that the former set of constraints in (\ref{x57a})
reduces to $\pi _{a_{2k+1}}^{(1)}\approx 0$. Thus, we observe that the
surface (\ref{94}--\ref{95}), (\ref{x57a}--\ref{x57abc}) results in a
trivial way from (\ref{94}--\ref{95}) by adding the canonical variables (\ref
{x57aa}--\ref{x57ab}), and demanding that their momenta vanish. Then, the
difference between an observable $F$ of the new irreducible theory and one
of the previous irreducible system, $\bar{F}$, is of the type $F-\bar{F}%
=\sum_{k=0}^{\Lambda }f^{a_{2k+1}}\pi _{a_{2k+1}}^{(1)}+\sum_{k=0}^{\Lambda
}g^{a_{2k+1}}\pi _{a_{2k+1}}^{(2)}+\sum_{k=1}^{\Gamma }h^{a_{2k}}\pi
_{a_{2k}}$, hence $F$ and $\bar{F}$ can be identified. Therefore, the
physical observables corresponding to the two irreducible systems coincide,
such that the supplementary constraints (\ref{x57a}--\ref{x57abc}) do not
afflict the previously established equivalence with the physical observables
of the original redundant theory. Consequently, we can replace the
Hamiltonian BRST-anti-BRST symmetry of the original reducible system with
that of the latter irreducible theory, and similarly with regard to the
associated path integrals.

>From now on, the Hamiltonian BRST-anti-BRST quantization of the irreducible
theory follows the standard lines. Defining a canonical action for $\left(
s_{a}\right) _{a=1,2}$ in the usual way as $s_{a}\bullet =\left[ \bullet
,\Omega _{a}\right] $, with $\left( \Omega _{a}\right) _{a=1,2}$ the BRST,
respectively, anti-BRST charge, the nilpotency and anticommutativity of $%
s_{a}$ imply that $\Omega _{a}$ should satisfy the equations 
\begin{equation}
\left[ \Omega _{a},\Omega _{b}\right] =0,\;a,b=1,2,  \label{x57xa}
\end{equation}
where $bingh\left( \Omega _{1}\right) =(1,0)$, $bingh\left( \Omega
_{2}\right) =(0,1)$. The existence of the solution to equations (\ref{x57xa}%
) is then guaranteed by the biacyclicity of the irreducible Koszul-Tate
bicomplex at positive total resolution degrees. Once we have constructed the
total BRST-anti-BRST charge, $\Omega ^{T}=\Omega _{1}+\Omega _{2}$, we pass
to the construction of the BRST-anti-BRST-invariant Hamiltonian, $%
H_{B}^{T}=H^{\prime }+{\rm ``more"}$, that satisfies $bingh\left(
H_{B}^{T}\right) =(0,0)$, $\left[ H_{B}^{T},\Omega ^{T}\right] =0$, where $%
H^{\prime }$ denotes the first-class Hamiltonian with respect to the
constraints (\ref{94}--\ref{95}) and (\ref{x57a}--\ref{x57abc}). In order to
fix the gauge, we have to choose a gauge-fixing fermion $K^{T}$ that
implements some irreducible gauge conditions and which is taken such that $%
\left( s_{1}+s_{2}\right) K^{T}=\left[ K^{T},\Omega ^{T}\right] $ contains
only terms of new ghost bigrading of the form $\left( k,k\right) $. It has
been shown in \cite{5} that under these circumstances the gauge-fixed
Hamiltonian $H_{K}^{T}=H_{B}^{T}+\left[ K^{T},\Omega ^{T}\right] $ is both
BRST and anti-BRST invariant and, moreover, produces a correct gauge-fixed
action $S_{K}^{T}$ with respect to the irreducible theory. In this way, we
showed how a reducible first-class Hamiltonian system can be approached
along an irreducible BRST-anti-BRST procedure. This completes our treatment.

\section{Example:\ the Freedman-Townsend model}

The starting point is the Lagrangian action of the Freedman-Townsend model 
\cite{ft} 
\begin{equation}
S_{0}^{L}\left[ A_{\mu }^{a},B_{a}^{\mu \nu }\right] =\frac{1}{2}\int
d^{4}x\left( -B_{a}^{\mu \nu }F_{\mu \nu }^{a}+A_{\mu }^{a}A_{a}^{\mu
}\right) ,  \label{ft1}
\end{equation}
where $B_{a}^{\mu \nu }$ denotes a set of antisymmetric tensor fields, and
the field strength of $A_{\mu }^{a}$ reads as $F_{\mu \nu }^{a}=\partial
_{\mu }A_{\nu }^{a}-\partial _{\nu }A_{\mu }^{a}-f_{\;\;bc}^{a}A_{\mu
}^{b}A_{\nu }^{c}$. The canonical analysis of this theory outputs the
constraints 
\begin{equation}
\Phi _{i}^{(1)a}\equiv \epsilon _{0ijk}\pi ^{jka}\approx 0,\;\Phi
_{i}^{(2)a}\equiv \frac{1}{2}\epsilon _{0ijk}\left( F^{jka}-\left( D^{\left[
j\right. }\right) _{\;\;b}^{a}\pi ^{\left. k\right] 0b}\right) \approx 0,
\label{ft2}
\end{equation}
\begin{equation}
\chi _{i}^{(1)a}\equiv \pi _{0i}^{a}\approx 0,\;\chi _{i}^{(2)a}\equiv \pi
_{i}^{a}+B_{0i}^{a}\approx 0,\;\chi _{a}^{(1)}\equiv \pi _{a}^{0}\approx 0,
\label{ft3}
\end{equation}
\begin{equation}
\chi _{a}^{(2)}\equiv A_{a}^{0}+f_{\;\;ab}^{c}B_{c}^{0i}\pi _{0i}^{b}+\left(
D_{i}\right) _{a}^{\;\;b}\pi _{b}^{i}\approx 0,  \label{ft4}
\end{equation}
and the first-class Hamiltonian 
\begin{eqnarray}
&&H=\int d^{3}x\left( \frac{1}{2}B_{a}^{ij}\left( F_{ij}^{a}-\left(
D_{\left[ i\right. }\right) _{\;\;b}^{a}\pi _{\left. j\right] 0}^{b}\right) -%
\frac{1}{2}A_{\mu }^{a}A_{a}^{\mu }-\right.   \nonumber  \label{ft5} \\
&&\left. A_{0}^{a}\left( \left( D_{i}\right) _{a}^{\;\;b}\pi
_{b}^{i}+f_{\;\;ab}^{c}B_{c}^{0i}\pi _{0i}^{b}\right) -A_{a}^{i}\left( \pi
_{0i}^{a}-\partial _{i}\pi _{0}^{a}\right) \right) .
\end{eqnarray}
The symbol $\left[ ij\right] $ appearing in (\ref{ft5}) signifies the
antisymmetry with respect to the indices between brackets. In the above, the
notations $\pi _{a}^{\mu }$ and $\pi _{\mu \nu }^{a}$ denote the momenta
respectively conjugated in the Poisson bracket to the fields $A_{\mu }^{a}$
and $B_{a}^{\mu \nu }$, while the covariant derivatives are defined by $%
\left( D_{i}\right) _{\;\;b}^{a}=\delta _{\;\;b}^{a}\partial
_{i}+f_{\;\;bc}^{a}A_{i}^{c}$ and $\left( D_{i}\right) _{b}^{\;\;a}=\delta
_{b}^{\;\;a}\partial _{i}-f_{\;\;bc}^{a}A_{i}^{c}$. By computing the Poisson
brackets between the constraint functions (\ref{ft2}--\ref{ft4}) we find
that (\ref{ft2}) are first-class and (\ref{ft3}--\ref{ft4}) second-class. In
addition, the functions $\Phi _{i}^{(2)a}$ from (\ref{ft2}) are on-shell
first-stage reducible 
\begin{equation}
\left( \left( D^{i}\right) _{\;\;b}^{a}+f_{\;\;bc}^{a}\pi ^{0ic}\right) \Phi
_{i}^{(2)b}=-\epsilon ^{0ijk}f_{\;\;bc}^{a}\chi _{i}^{(1)b}\left(
D_{j}\right) _{\;\;d}^{c}\chi _{k}^{(1)d}\approx 0.  \label{ft6}
\end{equation}
In order to deal with the Hamiltonian BRST-anti-BRST formalism, it is useful
to eliminate the second-class constraints with the help of the Dirac bracket 
\cite{Dirac} built with respect to themselves. By passing to the Dirac
bracket, the constraints (\ref{ft3}--\ref{ft4}) can be regarded as strong
equalities with the help of which we can express $A_{0}^{a}$, $\pi _{a}^{0}$%
, $\pi _{i}^{a}$ and $\pi _{0i}^{a}$ in terms of the remaining fields and
momenta, such that the independent `co-ordinates' of the reduced phase-space
are $A_{i}^{a}$, $B_{a}^{0i}$, $B_{a}^{ij}$ and $\pi _{ij}^{a}$. The
non-vanishing Dirac brackets among the independent components are expressed
by 
\begin{equation}
\left[ B_{a}^{0i}\left( x\right) ,A_{j}^{b}\left( y\right) \right]
_{x^{0}=y^{0}}^{*}=\delta _{a}^{\;\;b}\delta _{\;\;j}^{i}\delta ^{3}\left( 
{\bf x}-{\bf y}\right) ,  \label{ft7}
\end{equation}
\begin{equation}
\left[ B_{a}^{ij}\left( x\right) ,\pi _{kl}^{b}\left( y\right) \right]
_{x^{0}=y^{0}}^{*}=\frac{1}{2}\delta _{a}^{\;\;b}\delta _{\;\;k}^{\left[
i\right. }\delta _{\;\;l}^{\left. j\right] }\delta ^{3}\left( {\bf x}-{\bf y}%
\right) .  \label{ft8}
\end{equation}
In terms of the independent fields, the first-class constraints and
first-class Hamiltonian take the form 
\begin{equation}
\gamma _{i}^{(1)a}\equiv \epsilon _{0ijk}\pi ^{jka}\approx
0,\;G_{i}^{(2)a}\equiv \frac{1}{2}\epsilon _{0ijk}F^{jka}\approx 0,
\label{ft9}
\end{equation}
\begin{equation}
\tilde{H}=\frac{1}{2}\int d^{3}x\left(
B_{a}^{ij}F_{ij}^{a}-A_{i}^{a}A_{a}^{i}+\left( \left( D^{i}\right)
_{\;\;b}^{a}B_{0i}^{b}\right) \left( D_{j}\right)
_{a}^{\;\;c}B_{c}^{0j}\right) \equiv \int d^{3}x\,\tilde{h},  \label{ft10}
\end{equation}
while the reducibility relations 
\begin{equation}
\left( D^{i}\right) _{\;\;a}^{b}G_{i}^{(2)a}=0,  \label{ft11}
\end{equation}
hold off-shell in this case. Moreover, the first-class constraints (\ref{ft9}%
) remain Abelian in terms of the Dirac bracket. In the following we work
with the theory based on the reducible first-class constraints (\ref{ft9})
and on the first-class Hamiltonian (\ref{ft10}), in the context of the Dirac
bracket defined by (\ref{ft7}--\ref{ft8}).

In conclusion, this is an example of first-stage reducible theory ($L=1$),
with $G_{a_{0}}\rightarrow G_{i}^{(2)a}$ and $Z_{\;\;a_{1}}^{a_{0}}%
\rightarrow \left( D^{i}\right) _{\;\;a}^{b}$. Acting like in subsection
3.1, we add the bosonic scalar pairs $\left( \varphi _{b},\pi ^{b}\right) $
(that play the role of the variables $\left( y^{a_{1}},\pi _{a_{1}}\right) $%
) and take the matrix $A_{a_{0}}^{\;\;a_{1}}$ of the form 
\begin{equation}
A_{a_{0}}^{\;\;a_{1}}\rightarrow -\left( D_{i}\right) _{\;\;b}^{a},
\label{ft11a}
\end{equation}
such that condition (\ref{43}) is indeed satisfied. Consequently, from (\ref
{54}) and (\ref{ft11a}) we deduce the irreducible first-class constraints
associated with $G_{i}^{(2)a}\approx 0$ of the type 
\begin{equation}
\gamma _{a_{0}}\approx 0\rightarrow \gamma _{i}^{(2)a}\equiv \frac{1}{2}%
\epsilon _{0ijk}F^{jka}-\left( D_{i}\right) _{\;\;b}^{a}\pi ^{b}\approx 0.
\label{ft11b}
\end{equation}
Now, it is easy to see that the constraint functions in (\ref{ft11b}) and $%
\gamma _{i}^{(1)a}$ from (\ref{ft9}) are also first-class and irreducible.
In addition, as we mentioned at the end of the subsection 4.2 (see relations
(\ref{x57aa}) and (\ref{x57a})), we enlarge the phase-space with the
supplementary bosonic scalar pairs 
\begin{equation}
\left( y^{(1)a_{1}},\pi _{a_{1}}^{(1)}\right) \rightarrow \left( \varphi
_{a}^{(1)},\pi ^{(1)a}\right) ,\;\left( y^{(2)a_{1}},\pi
_{a_{1}}^{(2)}\right) \rightarrow \left( \varphi _{a}^{(2)},\pi
^{(2)a}\right) ,  \label{ft11c}
\end{equation}
subject to the constraints 
\begin{equation}
\gamma _{a_{1}}^{(1)}\approx 0\rightarrow \gamma ^{(1)a}\equiv \pi ^{a}-\pi
^{(1)a}\approx 0,\;\gamma _{a_{1}}^{(2)}\approx 0\rightarrow \gamma
^{(2)a}\equiv -\pi ^{(2)a}\approx 0,  \label{ft11d}
\end{equation}
whose presence does not in any way harm the irreducible theory, and,
moreover, is helpful at deriving a covariant form of the path integral
corresponding to the irreducible model. So far, we derived an irreducible
system for the Freedman-Townsend model, based on the irreducible first-class
constraints 
\begin{equation}
\gamma _{i}^{(1)a}\equiv \epsilon _{0ijk}\pi ^{jka}\approx 0,\;\gamma
_{i}^{(2)a}\equiv \frac{1}{2}\epsilon _{0ijk}F^{jka}-\left( D_{i}\right)
_{\;\;b}^{a}\pi ^{b}\approx 0,  \label{ft12}
\end{equation}
\begin{equation}
\gamma ^{(1)a}\equiv \pi ^{a}-\pi ^{(1)a}\approx 0,\;\gamma ^{(2)a}\equiv
-\pi ^{(2)a}\approx 0.  \label{ft13}
\end{equation}
The first-class Hamiltonian with respect to the above constraints can be
chosen of the type 
\begin{eqnarray}
&&H^{\prime }\equiv \int d^{3}x\,h^{\prime }=\int d^{3}x\left( \frac{1}{2}%
B_{a}^{ij}\left( F_{ij}^{a}+\epsilon _{0ijk}\left( D^{k}\right)
_{\;\;b}^{a}\pi ^{b}\right) -\right.   \nonumber  \label{ft14} \\
&&\frac{1}{2}A_{i}^{a}A_{a}^{i}+\varphi _{a}\pi ^{(2)a}-\varphi
_{a}^{(2)}\left( D_{i}\right) _{\;\;b}^{a}\left( D^{i}\right)
_{\;\;c}^{b}\pi ^{c}+  \nonumber \\
&&\left. \frac{1}{2}\left( \left( D_{i}\right)
_{a}^{\;\;b}B_{b}^{0i}-f_{\;\;ab}^{c}\left( \varphi _{c}\pi ^{b}+\varphi
_{c}^{(1)}\pi ^{b}+\varphi _{c}^{(2)}\pi ^{(2)b}\right) \right) ^{2}\right) ,
\end{eqnarray}
where we employed the notation 
\begin{eqnarray}
&&\left( \left( D_{i}\right) _{a}^{\;\;b}B_{b}^{0i}-f_{\;\;ab}^{c}\left(
\varphi _{c}\pi ^{b}+\varphi _{c}^{(1)}\pi ^{b}+\varphi _{c}^{(2)}\pi
^{(2)b}\right) \right) ^{2}\equiv   \nonumber  \label{ft15} \\
&&\left( \left( D_{i}\right) _{a}^{\;\;b}B_{b}^{0i}-f_{\;\;ab}^{c}\left(
\varphi _{c}\pi ^{b}+\varphi _{c}^{(1)}\pi ^{b}+\varphi _{c}^{(2)}\pi
^{(2)b}\right) \right) \times   \nonumber \\
&&\left( \left( D^{j}\right) _{\;\;d}^{a}B_{0j}^{d}-f_{\;\;de}^{a}\left( \pi
^{d}\varphi ^{e}+\pi ^{d}\varphi ^{(1)e}+\pi ^{(2)d}\varphi ^{(2)e}\right)
\right) .
\end{eqnarray}
The irreducible first-class constraints are Abelian, while the remaining
gauge algebra relations read as 
\begin{eqnarray}
&&\left[ \gamma _{i}^{(1)a},H^{\prime }\right] ^{*}=-\gamma
_{i}^{(2)a},\;\left[ \gamma _{i}^{(2)a},H^{\prime }\right] ^{*}=-\left(
D_{i}\right) _{\;\;b}^{a}\gamma ^{(2)b}+  \nonumber  \label{ft16} \\
&&f_{\;\;bc}^{a}\left( \left( D^{j}\right)
_{\;\;d}^{b}B_{0j}^{d}-f_{\;\;de}^{b}\left( \pi ^{d}\varphi ^{e}+\pi
^{d}\varphi ^{(1)e}+\pi ^{(2)d}\varphi ^{(2)e}\right) \right) \gamma
_{i}^{(2)c},
\end{eqnarray}
\begin{eqnarray}
&&\left[ \gamma ^{(1)a},H^{\prime }\right] ^{*}=\gamma ^{(2)a},\;\left[
\gamma ^{(2)a},H^{\prime }\right] ^{*}=\left( D^{i}\right)
_{\;\;b}^{a}\gamma _{i}^{(2)b}+  \nonumber  \label{ft17} \\
&&f_{\;\;bc}^{a}\left( \left( D^{j}\right)
_{\;\;d}^{b}B_{0j}^{d}-f_{\;\;de}^{b}\left( \pi ^{d}\varphi ^{e}+\pi
^{d}\varphi ^{(1)e}+\pi ^{(2)d}\varphi ^{(2)e}\right) \right) \gamma ^{(2)c}.
\end{eqnarray}
As will be emphasized, the gauge algebra (\ref{ft16}--\ref{ft17}) ensures
the Lorentz covariance of the irreducible approach.

Next, we determine the path integral of the irreducible first-class theory
described by (\ref{ft12}--\ref{ft14}), associated with the Freedman-Townsend
model, in the context of the Hamiltonian BRST-anti-BRST formalism. In view
of this, we introduce the generators of the BRST-anti-BRST bicomplex 
\begin{equation}
\left( \stackrel{(-1,0)}{\cal P}_{1i}^{(1)a},\stackrel{(0,-1)}{\cal P}%
_{2i}^{(1)a},\stackrel{(-1,0)}{\cal P}_{1}^{(1)a},\stackrel{(0,-1)}{\cal P}%
_{2}^{(1)a}\right) ,  \label{ft18}
\end{equation}
\begin{equation}
\left( \stackrel{(-1,0)}{\cal P}_{1i}^{(2)a},\stackrel{(0,-1)}{\cal P}%
_{2i}^{(2)a},\stackrel{(-1,0)}{\cal P}_{1}^{(2)a},\stackrel{(0,-1)}{\cal P}%
_{2}^{(2)a}\right) ,  \label{ft18a}
\end{equation}
\begin{equation}
\left( \stackrel{(-1,-1)}{\lambda }_{i}^{(1)a},\stackrel{(-1,-1)}{\lambda }%
^{(1)a},\stackrel{(-1,-1)}{\lambda }_{i}^{(2)a},\stackrel{(-1,-1)}{\lambda }%
^{(2)a}\right) ,  \label{ft19}
\end{equation}
\begin{equation}
\left( \stackrel{(1,0)}{\eta }_{1a}^{(1)i},\stackrel{(0,1)}{\eta }%
_{2a}^{(1)i},\stackrel{(1,0)}{\eta }_{1a}^{(1)},\stackrel{(0,1)}{\eta }%
_{2a}^{(1)}\right) ,  \label{ft20}
\end{equation}
\begin{equation}
\left( \stackrel{(1,0)}{\eta }_{1a}^{(2)i},\stackrel{(0,1)}{\eta }%
_{2a}^{(2)i},\stackrel{(1,0)}{\eta }_{1a}^{(2)},\stackrel{(0,1)}{\eta }%
_{2a}^{(2)}\right) ,  \label{ft20a}
\end{equation}
\begin{equation}
\left( \stackrel{(1,1)}{Q}_{a}^{(1)i},\stackrel{(1,1)}{Q}_{a}^{(1)},%
\stackrel{(1,1)}{Q}_{a}^{(2)i},\stackrel{(1,1)}{Q}_{a}^{(2)}\right) ,
\label{ft21}
\end{equation}
graded according to the new ghost bidegree. The total BRST-anti-BRST charge
has the form 
\begin{eqnarray}
&&\Omega ^{T}=\int d^{3}x\left( \gamma _{i}^{(1)a}\left( \eta
_{1a}^{(1)i}+\eta _{2a}^{(1)i}\right) +\gamma _{i}^{(2)a}\left( \eta
_{1a}^{(2)i}+\eta _{2a}^{(2)i}\right) +\gamma ^{(1)a}\left( \eta
_{1a}^{(1)}+\eta _{2a}^{(1)}\right) +\right.   \nonumber  \label{ft22} \\
&&\gamma ^{(2)a}\left( \eta _{1a}^{(2)}+\eta _{2a}^{(2)}\right)
+Q_{a}^{(1)i}\left( {\cal P}_{1i}^{(1)a}-{\cal P}_{2i}^{(1)a}\right)
+Q_{a}^{(2)i}\left( {\cal P}_{1i}^{(2)a}-{\cal P}_{2i}^{(2)a}\right) + 
\nonumber \\
&&\left. Q_{a}^{(1)}\left( {\cal P}_{1}^{(1)a}-{\cal P}_{2}^{(1)a}\right)
+Q_{a}^{(2)}\left( {\cal P}_{1}^{(2)a}-{\cal P}_{2}^{(2)a}\right) \right) .
\end{eqnarray}
The BRST-anti-BRST-invariant Hamiltonian corresponding to the first-class
Hamiltonian (\ref{ft14}) is given by 
\begin{eqnarray}
&&H_{B}^{T}=H^{\prime }+\int d^{3}x\left( \eta _{1a}^{(1)i}{\cal P}%
_{1i}^{(2)a}+\eta _{2a}^{(1)i}{\cal P}_{2i}^{(2)a}+Q_{a}^{(1)i}\lambda
_{i}^{(2)a}-\eta _{1a}^{(1)}{\cal P}_{1}^{(2)a}-\eta _{2a}^{(1)}{\cal P}%
_{2}^{(2)a}-\right.   \nonumber  \label{ft23} \\
&&Q_{a}^{(1)}\lambda ^{(2)a}+\eta _{1a}^{(2)i}\left( D_{i}\right)
_{\;\;b}^{a}{\cal P}_{1}^{(2)b}+\eta _{2a}^{(2)i}\left( D_{i}\right)
_{\;\;b}^{a}{\cal P}_{2}^{(2)b}+Q_{a}^{(2)i}\left( D_{i}\right)
_{\;\;b}^{a}\lambda ^{(2)b}-  \nonumber \\
&&\eta _{1a}^{(2)}\left( D^{i}\right) _{\;\;b}^{a}{\cal P}_{1i}^{(2)b}-\eta
_{2a}^{(2)}\left( D^{i}\right) _{\;\;b}^{a}{\cal P}_{2i}^{(2)b}-Q_{a}^{(2)}%
\left( D^{i}\right) _{\;\;b}^{a}\lambda _{i}^{(2)b}-  \nonumber \\
&&f_{\;\;ab}^{c}\left( \eta _{1c}^{(2)}{\cal P}_{1}^{(2)b}+\eta _{2c}^{(2)}%
{\cal P}_{2}^{(2)b}+Q_{c}^{(2)}\lambda ^{(2)b}+\eta _{1c}^{(2)i}{\cal P}%
_{1i}^{(2)b}+\eta _{2c}^{(2)i}{\cal P}_{2i}^{(2)b}+Q_{c}^{(2)i}\lambda
_{i}^{(2)b}\right) \times   \nonumber \\
&&\times \left( \left( D^{j}\right)
_{\;\;d}^{a}B_{0j}^{d}-f_{\;\;de}^{a}\left( \pi ^{d}\varphi ^{e}+\pi
^{d}\varphi ^{(1)e}+\pi ^{(2)d}\varphi ^{(2)e}\right) \right) +  \nonumber \\
&&\frac{1}{2}\left( f_{\;\;ab}^{c}\left( \eta _{1c}^{(2)}{\cal P}%
_{1}^{(2)b}+\eta _{2c}^{(2)}{\cal P}_{2}^{(2)b}+Q_{c}^{(2)}\lambda
^{(2)b}+\right. \right.   \nonumber \\
&&\left. \left. \left. \eta _{1c}^{(2)i}{\cal P}_{1i}^{(2)b}+\eta
_{2c}^{(2)i}{\cal P}_{2i}^{(2)b}+Q_{c}^{(2)i}\lambda _{i}^{(2)b}\right)
\right) ^{2}\right) ,
\end{eqnarray}
where we use the notation 
\begin{eqnarray}
&&\left( f_{\;\;ab}^{c}\left( \eta _{1c}^{(2)}{\cal P}_{1}^{(2)b}+\eta
_{2c}^{(2)}{\cal P}_{2}^{(2)b}+Q_{c}^{(2)}\lambda ^{(2)b}+\right. \right.  
\nonumber  \label{ft24} \\
&&\left. \left. \eta _{1c}^{(2)i}{\cal P}_{1i}^{(2)b}+\eta _{2c}^{(2)i}{\cal %
P}_{2i}^{(2)b}+Q_{c}^{(2)i}\lambda _{i}^{(2)b}\right) \right) ^{2}\equiv  
\nonumber \\
&&-f_{\;\;ab}^{c}\left( \eta _{1c}^{(2)}{\cal P}_{1}^{(2)b}+\eta _{2c}^{(2)}%
{\cal P}_{2}^{(2)b}+Q_{c}^{(2)}\lambda ^{(2)b}+\right.   \nonumber \\
&&\left. \eta _{1c}^{(2)i}{\cal P}_{1i}^{(2)b}+\eta _{2c}^{(2)i}{\cal P}%
_{2i}^{(2)b}+Q_{c}^{(2)i}\lambda _{i}^{(2)b}\right) \times   \nonumber \\
&&\times f_{\;\;de}^{a}\left( \eta _{1}^{(2)d}{\cal P}_{1}^{(2)e}+\eta
_{2}^{(2)d}{\cal P}_{2}^{(2)e}+Q^{(2)d}\lambda ^{(2)e}+\right.   \nonumber \\
&&\left. \eta _{1}^{(2)di}{\cal P}_{1i}^{(2)e}+\eta _{2}^{(2)di}{\cal P}%
_{2i}^{(2)e}+Q^{(2)di}\lambda _{i}^{(2)e}\right) .
\end{eqnarray}
In order to obtain the path integral of the irreducible model, we work with
the gauge-fixing fermion 
\begin{eqnarray}
&&K^{T}=\int d^{3}x\left( {\cal P}_{1i}^{(1)a}\left( \epsilon
^{0ijk}\partial _{j}B_{0ka}+\partial ^{i}\varphi _{a}^{(1)}\right) -\frac{1}{%
2}{\cal P}_{1}^{(1)a}\epsilon ^{0ijk}\partial _{i}B_{jka}+\lambda
_{i}^{(1)a}\partial ^{i}\eta _{2a}^{(1)}+\right.   \nonumber  \label{ft25} \\
&&\left. \frac{1}{2}\partial _{\left[ i\right. }\lambda _{\left. j\right]
}^{(1)a}\left( D^{\left[ i\right. }\right) _{a}^{\;\;b}\eta _{2b}^{(2)\left.
j\right] }+f_{\;\;bc}^{a}\epsilon ^{0ijk}\pi ^{b}\eta _{2ia}^{(2)}\partial
_{j}\lambda _{k}^{(1)c}-\lambda ^{(1)a}\partial _{i}\eta _{2a}^{(1)i}\right)
.
\end{eqnarray}
If we compute the path integral with the help of the above gauge-fixing
fermion, we find that its exponent contains a quadratic term, namely, 
\begin{eqnarray}
&&\exp \left( i\int d^{4}x\left( -\frac{1}{2}\left( \left( D_{i}\right)
_{a}^{\;\;b}B_{b}^{0i}-f_{\;\;ab}^{c}\left( \varphi _{c}\pi ^{b}+\varphi
_{c}^{(1)}\pi ^{b}+\varphi _{c}^{(2)}\pi ^{(2)b}\right) -\right. \right.
\right.   \nonumber  \label{ft26} \\
&&f_{\;\;ab}^{c}\left( \eta _{1c}^{(2)}{\cal P}_{1}^{(2)b}+\eta _{2c}^{(2)}%
{\cal P}_{2}^{(2)b}+Q_{c}^{(2)}\lambda ^{(2)b}+\right.   \nonumber \\
&&\left. \left. \left. \left. \eta _{1c}^{(2)i}{\cal P}_{1i}^{(2)b}+\eta
_{2c}^{(2)i}{\cal P}_{2i}^{(2)b}+Q_{c}^{(2)i}\lambda _{i}^{(2)b}\right)
\right) ^{2}\right) \right) ,
\end{eqnarray}
that can be equivalently written in a linearized form by means of
introducing a new field through the Gaussian average 
\begin{eqnarray}
&&\int {\cal D}H_{0}^{a}\exp \left( i\int d^{4}x\left( \frac{1}{2}%
H_{0}^{a}H_{a}^{0}-H_{0}^{a}\left( \left( D_{i}\right)
_{a}^{\;\;b}B_{b}^{0i}-\right. \right. \right.   \nonumber  \label{ft27} \\
&&f_{\;\;ab}^{c}\left( \varphi _{c}\pi ^{b}+\varphi _{c}^{(1)}\pi
^{b}+\varphi _{c}^{(2)}\pi ^{(2)b}\right) -f_{\;\;ab}^{c}\left( \eta
_{1c}^{(2)}{\cal P}_{1}^{(2)b}+\right.   \nonumber \\
&&\left. \left. \left. \left. \eta _{2c}^{(2)}{\cal P}%
_{2}^{(2)b}+Q_{c}^{(2)}\lambda ^{(2)b}+\eta _{1c}^{(2)i}{\cal P}%
_{1i}^{(2)b}+\eta _{2c}^{(2)i}{\cal P}_{2i}^{(2)b}+Q_{c}^{(2)i}\lambda
_{i}^{(2)b}\right) \right) \right) \right) .
\end{eqnarray}
Eliminating some of the auxiliary variables from the resulting gauge-fixed
action on their equations of motion, we finally arrive at 
\begin{eqnarray}
&&Z_{K^{T}}=\int {\cal D}A_{\mu }^{a}{\cal D}B_{a}^{\mu \nu }{\cal D}\varphi
_{a}^{(1)}{\cal D}b_{\mu }^{a}{\cal D}Q_{a}^{(2)\mu }{\cal D}\lambda _{\mu
}^{(1)a}\times   \nonumber  \label{ft28} \\
&&\times {\cal D}\eta _{1a}^{(2)\mu }{\cal DP}_{1\mu }^{(1)a}{\cal D}\eta
_{2a}^{(2)\mu }{\cal DP}_{2\mu }^{(1)a}\exp \left( iS_{K}^{T}\right) ,
\end{eqnarray}
where 
\begin{eqnarray}
&&S_{K}^{T}=S_{0}^{L}\left[ A_{\mu }^{a},B_{a}^{\mu \nu }\right] +\int
d^{4}x\left( b_{\mu }^{a}\left( \frac{1}{2}\epsilon ^{\mu \nu \lambda \rho
}\partial _{\nu }B_{\lambda \rho a}+\partial ^{\mu }\varphi
_{a}^{(1)}\right) -\right.   \nonumber  \label{ft29} \\
&&\frac{1}{2}\partial _{\left[ \mu \right. }\lambda _{\left. \nu \right]
}^{(1)a}\left( D^{\left[ \mu \right. }\right) _{a}^{\;\;b}Q_{b}^{(2)\left.
\nu \right] }-\left( \partial ^{\mu }\lambda _{\mu }^{(1)a}\right) \left(
D_{\nu }\right) _{a}^{\;\;b}Q_{b}^{(2)\nu }+  \nonumber \\
&&\frac{1}{2}\partial _{\left[ \mu \right. }{\cal P}_{1\left. \nu \right]
}^{(1)a}\left( D^{\left[ \mu \right. }\right) _{a}^{\;\;b}\eta
_{1b}^{(2)\left. \nu \right] }+\left( \partial ^{\mu }{\cal P}_{1\mu
}^{(1)a}\right) \left( D_{\nu }\right) _{a}^{\;\;b}\eta _{1b}^{(2)\nu }+ 
\nonumber \\
&&\left. \frac{1}{2}\partial _{\left[ \mu \right. }{\cal P}_{2\left. \nu
\right] }^{(1)a}\left( D^{\left[ \mu \right. }\right) _{a}^{\;\;b}\eta
_{2b}^{(2)\left. \nu \right] }+\left( \partial ^{\mu }{\cal P}_{2\mu
}^{(1)a}\right) \left( D_{\nu }\right) _{a}^{\;\;b}\eta _{2b}^{(2)\nu
}\right) ,
\end{eqnarray}
with $S_{0}^{L}\left[ A_{\mu }^{a},B_{a}^{\mu \nu }\right] $ the original
action (\ref{ft1}). In inferring the above covariant form of the path
integral we performed the identifications 
\begin{equation}
A_{\mu }^{a}\equiv \left( H_{0}^{a},A_{i}^{a}\right) ,\;b_{\mu }^{a}\equiv
\left( \pi ^{(1)a},\epsilon _{0ijk}\pi ^{jka}\right) ,  \label{ft30}
\end{equation}
\begin{equation}
Q_{a}^{(2)\mu }\equiv \left( Q_{a}^{(2)},Q_{a}^{(2)i}\right) ,\;\lambda
_{\mu }^{(1)a}\equiv \left( -\lambda ^{(1)a},\lambda _{i}^{(1)a}\right) ,
\label{ft30a}
\end{equation}
\begin{equation}
\eta _{1a}^{(2)\mu }\equiv \left( -\eta _{1a}^{(2)},\eta _{1a}^{(2)i}\right)
,\;{\cal P}_{1\mu }^{(1)a}\equiv \left( {\cal P}_{1}^{(1)a},{\cal P}%
_{1i}^{(1)a}\right) ,  \label{ft30b}
\end{equation}
\begin{equation}
\eta _{2a}^{(2)\mu }\equiv \left( -\eta _{2a}^{(2)},\eta _{2a}^{(2)i}\right)
,\;{\cal P}_{2\mu }^{(1)a}\equiv \left( {\cal P}_{2}^{(1)a},{\cal P}%
_{2i}^{(1)a}\right) ,  \label{ft31}
\end{equation}
and, in addition, we adopted the notation 
\begin{equation}
\left( D_{0}\right) _{a}^{\;\;b}=\delta _{a}^{\;\;b}\partial
_{0}-f_{\;\;ac}^{b}H_{0}^{c}.  \label{ft32}
\end{equation}
This ends the irreducible BRST-anti-BRST Hamiltonian treatment of the model
under study.

\section{Conclusion}

To conclude with, in this paper we have exposed an irreducible Hamiltonian
BRST-anti-BRST method for quantizing reducible first-class systems. The key
point of our approach consists in the construction of an irreducible
Koszul-Tate bicomplex associated with that of the reducible theory, that
reveals some irreducible first-class constraints. Moreover, the physical
observables of the starting reducible system and of the resulting
irreducible one coincide. This result together with the basic equations of
the Hamiltonian BRST-anti-BRST method allow the replacement of the
Hamiltonian BRST-anti-BRST quantization of the initial reducible system with
that of the irreducible theory. The existence of the canonical generator of
the irreducible BRST-anti-BRST symmetry is provided by the biacyclicity of
the irreducible Koszul-Tate bicomplex, while the gauge-fixing procedure is
facilitated by the enlargement of the phase-space with the canonical pairs
of the type $\left( y,\pi \right) $. The theoretical part of the paper has
been exemplified on the Freedman-Townsend model.

\section*{Acknowledgment}

This work has been supported by a Romanian National Council for Academic
Scientific Research (CNCSIS) grant.

\end{document}